\documentstyle[multicol,epsfig,epsf,aps,prl]{revtex}
\begin{document}
\topskip 20mm
\draft
\tightenlines
\title{ Level-statistics of Anderson model of disordered systems: Connection to Brownian Ensembles}
\author{Pragya Shukla$^{*}$}
\address{ Department of Physics,Indian Institute of Technology, Kharagpur-721302, India}
\onecolumn
\date{\today}
\maketitle
\widetext
\begin{abstract}

	We find that the statistics of levels undergoing 
metal-insulator transition in systems with  Gaussian 
disorder and non-interacting electrons behaves in a way similar to that 
of the single parametric Brownian ensembles \cite{dy}. 
The latter appear during a Poisson $\rightarrow$ Wigner-Dyson transition, 
driven by a random perturbation. The analogy  provides the analytical evidence 
for the single parameter scaling of the level-correlations in  
disordered systems as well as a tool to obtain them at the 
critical point for a wide range of disorders.

\end{abstract}

.

	The spectral correlations of a disordered system are very 
sensitive to the behavior of its eigenfunctions. The presence of disorder 
may cause localized waves in the system, implying lack of interaction between 
certain parts. This is reflected in the structure of the Hamiltonian matrix 
which is  sparse in the site representation. The degree of sparsity of the matrix 
is governed by various system parameters e.g. dimensionality, shape, size 
and boundary conditions of the system. The variation of the disorder-strength can lead to 
a metal-insulator transition (MIT), with eigenfunctions changing from a fully extended state 
(metal) to a strongly localized one (insulator) with partial localization in the critical 
region. The associated Hamiltonian also undergoes a transition, ({\it in effect only} due to 
variation of the relative strength of its elements), from a full matrix to a sparse or banded form 
and finally to a diagonal matrix. The statistical studies of levels for various degrees and types 
of disorders as well as system conditions require, therefore, analysis of different ensembles. Here 
the nature of the localization and its strength is reflected in the measure and the sparsity 
of the ensemble, respectively.  Our objective in this paper is to obtain a mathematical formulation 
for the level-correlations, common to a large class of system conditions (with Gaussian type 
randomness); the system information enters in the formulation 
through a parameter, basically 
a function of various system parameters influencing the localization.

	Recently it was shown  that the eigenvalue distributions
of various ensembles, with a multi-parametric Gaussian measure and independent 
matrix elements,  appear as 
non-equilibrium stages of a Brownian type diffusion process \cite{ps}. 
Here the eigenvalues evolve with respect to a single 
parameter which is a function of the distribution parameters of the 
ensemble. The parameter is therefore related to the complexity of the system 
represented by the ensemble and can be termed as the "complexity" parameter. 
The solution of the diffusion equation
for a given value of the complexity parameter gives the
distribution of the eigenvalues, and thereby their correlations, for the 
corresponding system. A similar diffusion equation is known to govern the 
evolution of the eigenvalues of Brownian Ensembles (BE) \cite{dy,me} 
and many of its solutions for various initial conditions have already been 
obtained \cite{ap}. The analogy can then be used to obtain the 
level-correlations for the Gaussian random matrix models of the 
disordered systems with non-interacting 
electrons. The presence of interactions introduces a correlation between  
matrix elements of the ensemble representing the system; 
the details of this case are discussed elsewhere \cite{ps4}.

	The correlations in the single electron spectra of disordered 
metals are governed by a variety of parameters e.g the associated 
energy-ranges, degree of disorder, the dimensionality of the system etc. 
Here the two energy-scales, playing the dominant role, are the Thouless energy 
$E_c$ and the mean level spacing $\Delta$. The $E_c$ is given by the 
time-scale needed by the wave-packet to diffuse through the sample.   
In the diffusive (metallic) regime and for energy-scales $\delta E$ smaller 
then $E_c$, the spectral correlations are well-modeled by 
 Wigner-Dyson (WD) ensembles \cite{me}. 
For $\delta E >E_c$, 
the statistics deviate from the Wigner-Dyson case, however the deviations 
are negligible for sample size $L \rightarrow \infty$. 
In the localized (insulator) phase too, the correlations are 
energy-dependent but, in the limit $L\rightarrow \infty$, the 
levels are completely uncorrelated and their statistics can be modeled by 
the Poisson ensemble. However the statistics in the critical region near the 
metal-insulator transition (Anderson type) 
is different from both Wigner-Dyson as well as Poisson statistics and 
depends on various system dependent features\cite{ss,am}. 
Our study shows that the multi-parametric level-statistics in the critical 
region can be well-modeled by the single-parametric Brownian Ensembles.

	The paper is organized as follows. The section I contains a brief 
description of the simplest model of a disordered system using independent 
electron approximation and the equation governing the evolution of its eigenvalues 
due to change of disorder etc. 
The properties of the BEs useful for present study are given in 
section II. The section III deals with the determination of 
the single parameter $\Lambda$ governing the level-statistics during MIT 
using BE-analogy. It also provides an explanation, in terms of $\Lambda$, of 
some of the observed  features of the AE-statistics. In section IV, the AE-BE  
analogy is used to to obtain the analytical formulation of some of the unknown 
spectral fluctuations during MIT. The section V 
contains the details of the numerical comparison of the level-statistics 
of Anderson Hamiltonian with that of BEs and reconfirms our analytical results. 
The studies, during last decade, indicate the surprising success of  
power law random banded matrices (1D system) as a model for Anderson ensembles \cite{mir};
as discussed in section VI, the success can be explained within $\Lambda$ formulation 
of the level-statistics.

\section {The Multi-Dimensional Anderson Hamiltonian}

	The Anderson model for a disordered system is described by
 a $d$-dimensional disordered lattice, of size $L$, with a
  Hamiltonian $H= \sum_n \epsilon_n a_n^+ a_n -
\sum_{n\not=m} b_{mn} (a_n^+ a_m +a_n a_m^+)$ in tight-binding
approximation.
The site energies $\epsilon_n$, measured in units of the
overlap integral between adjacent sites, correspond to the random potential.
The hopping is  
assumed to connect only the $z$ nearest-neighbors (referred by $m$) 
 of each site. 
In the site representation, $H$
turns out to be a sparse matrix of size $N=L^d$ with diagonal
matrix elements as the site energies $H_{kk}=\epsilon_k$. The 
off-diagonals $H_{kl}$ describe the interaction between two sites $k$ and $l$; 
here $H_{kl}$ for two sites connected by hopping will be referred as 
hopping off-diagonal and the rest as non-hopping off-diagonals.  
 The level-statistics of $H$ can
therefore be studied by analyzing the properties of an ensemble of (i) sparse
real symmetric matrices, in presence of a time-reversal symmetry  and
(ii) sparse complex Hermitian matrices in absence of a time-reversal. 

	We consider an ensemble of Anderson Hamiltonians (later referred as Anderson 
ensemble) with a Gaussian type disorder.
The site-energies $H_{kk}=\epsilon_k$ are thus independent Gaussian 
distributions $\rho_{kk}(H_{kk})={\rm e}^{-(H_{kk}-b_{kk})^2/2h_{kk}}$ 
with variance $h_{kk}$ and mean $b_{kk}$. 
The hopping can be chosen to 
be isotropic or anisotropic, non-random or random (Gaussian). 
A general form of the probability density $\rho(H)\equiv \prod_{k,l;k\le l}
\rho_{kl}(H_{kl})$ of the ensemble, including all the above possibilities,  
 can therefore be given by

\begin{eqnarray}
 \rho (H,h,b)=C{\rm exp}[{-\sum_{s=1}^\beta \sum_{k\le l} (1/2 h_{kl;s}) 
 (H_{kl;s}-b_{kl;s})^2 }]
\end{eqnarray}
 with subscript $"s"$ of a variable referring to its components, 
$\beta$ as their 
total number ($\beta=1$ for real variable, $\beta=2$ for the complex one), 
$C$ as the normalization constant, $h$  as the set of the variances 
$h_{kl;s}=<H^2_{kl;s}>$
 and $b$ as the set of all mean values $<H_{kl;s}>=b_{kl;s}$.
As obvious, in the limit $h_{kl;1}, h_{kl;2} \rightarrow 0$,
eq.(1) corresponds to the
non-random nature of $H_{kl}$ (that is, $\rho_{kl}(H_{kl})=\delta(H_{kl}- b_{kl})$).
Note although the non-hopping off-diagonals in Anderson matrix 
always remain zero but the effective sparsity of the matrix changes 
due to change in relative strength of the diagonals and the 
hopping off-diagonals. 
Thus, in the insulator limit (with almost no overlap between site 
energies due to strong disorder), the matrix behaves effectively as 
a diagonal one, the diagonals being very large as compared to hopping 
off-diagonals. In the opposite 
limit of very weak disorder when an average diagonal is nearly of the same 
strength as an average off-diagonal, the statistical behavior of the 
matrix is same as that of a matrix taken from a Wigner-Dyson ensemble \cite{me}. 
The latter are       
the basis-invariant Gaussian ensembles of Hermitian type, with a 
same variance for almost all matrix elements. The statistical behavior 
of levels in the Wigner-Dyson ensembles depends only on their symmetry class and 
is therefore universal in nature. The three main universality classes 
are described by a parameter $\beta$, basically a measure of the degree of 
level-repulsion\cite{me}: 
(i) GOE with $\beta=1$, corresponding to time-reversal 
symmetry and integer 
angular momentum, (ii) GUE with $\beta=2$ and no time-reversal symmetry, 
(iii) GSE with $\beta=4$ and time-reversal symmetry but half integer 
angular momentum.

A variation of disorder and hopping rate changes the distribution parameters
 of the probability density $\rho(H)$ and 
thereby its statistical properties. 
Using Gaussian nature of $\rho$, it is easy to verify that under a
change of parameters $h_{kl} \rightarrow h_{kl}+\delta h_{kl}$ 
and $b_{kl} \rightarrow b_{kl}+\delta b_{kl}$, the matrix elements 
$H_{kl}$ undergo a 
diffusion dynamics along with a finite drift,

\begin{eqnarray}
& &\sum_{k\le l;s}\left[(2/\tilde g_{kl}) 
x_{kl;s}{\partial \rho\over\partial h_{kl;s}} - \gamma
  b_{kl;s} {\partial \rho\over\partial b_{kl;s}}\right]   = \nonumber \\
& &\sum_{kl;s} {\partial \over \partial H_{kl;s}}
\left[ {g_{kl}\over 2} {\partial \over \partial H_{kl;s}} +
 \gamma H_{kl;s} \right] \rho
\end{eqnarray}
where $x_{kl;s} \equiv 1-\gamma {\tilde g_{kl}} h_{kl;s}$ with 
${\tilde g_{kl}}=2-\delta_{kl}$ and $g_{kl}=1+\delta_{kl}$.
The $\gamma$ is an arbitrary parameter, giving  
the variance of the matrix elements at the end of the evolution\cite{ps}.
The above equation describes a multi parametric flow of matrix elements 
from an arbitrary initial condition, say $H_0$.  
However, as discussed in \cite{ps}, it is possible to define a "complexity"
parameter $Y$, a function of various distribution parameters $h_{kl;s}$ and 
$b_{kl;s}$, in terms of which the matrix elements undergo a single parametric 
diffusion,

\begin{eqnarray}
{\partial \rho\over\partial Y}   =
\sum_{kl;s} {\partial \over \partial H_{kl;s}}
\left[ {g_{kl}\over 2} {\partial \over \partial H_{kl;s}} +
 \gamma H_{kl;s} \right] \rho
\end{eqnarray}
with 
\begin{eqnarray}
Y= -{1\over 2 M \gamma} 
 {\rm ln}\left[ \prod_{k \le l}^{'}\prod_{s=1}^{\beta}
 |x_{kl;s}| \quad |b_{kl;s}|^2 \right] + C 
\end{eqnarray}
here $\prod'$ implies a product over non-zero $b_{kl;s}$ and $x_{kl;s}$.
Further, $C$ is a constant determined by the initial distribution, 
$M$ is the number of all non-zero parameters $x_{kl;s}$ and $b_{kl;s}$ 
and $\beta=1,2$ for Hamiltonians with and without time-reversal, 
respectively. 

	The solution of the eq.(3) gives the state $\rho(H,Y|H_0,Y_0)$  
of the flow at parameter $Y$, starting from an initial state $H_0$ 
with $Y=Y_0$. An integration over initial probability density result in 
the density given by eq.(1) in terms of $Y(h,b)$: 
$\rho(H,Y)=\int \rho(H,Y|H_0,Y_0)\rho(H_0,y_0) {\rm d}H_0$.  
The evolution reaches a steady state when 
${\partial \rho/\partial Y}\rightarrow 0$ with the ensemble $\rho(H)$ 
approaching the Wigner-Dyson limit, 
$\rho\propto {\rm e}^{-(\gamma/2) {\rm Tr}H^2}$.

	As implied by eq.(3), the variation of $\rho(H|H_0)$, 
depends on the changes of the parameters 
$h_{kl}$, $b_{kl}$ (for all $k,l$) only through a function $Y$.
This can be proved by considering a transformation of the 
$M$ non-zero variables of the sets $h$ and $b$   
to another set $\{Y,Y_2,..,Y_M\}$ of $M$ variables; 
$h_{kl}=h_{kl}(Y,Y_2,..,Y_M)$ and $b_{kl}=b_{kl}(Y,Y_2,..,Y_M)$.
As shown in \cite{ps,ps4}, it is possible to define $Y,Y_2,..Y_M$  such that 
the $M-1$ variables $Y_2,..,Y_M$ remain constant during the evolution 
of $\rho$ due to any change in sets $h,b$.
The statistics during the transition is therefore governed by $Y$ only. 
The choice of the $Y_2,..,Y_M$ depends on the system under consideration. 
For a transition preserving the lattice structure, these 
constants turn out to be the functions of the site-indices in the lattice. 
For example, the variances $h_{kl}$ in the Anderson ensemble are functions 
of the disorder as 
well as the site-indices $k$, $l$; $Y$ can then be identified as a 
function of disorder while $Y_j$ ($j>1$) as the functions of site indices.  
Further, as these constants do not appear explicitly in eq.(3), its solution and therefore 
the ensemble-statistics is independent of the specific values of the constants. 

 	The flow described by eq.(3) can start from  any initial state; 
the only constraint on the choice is that the parameters 
$Y_j$, $j>2$, for the initial ensemble should be same as those for the ensemble 
$\rho(H,h,b)$.  As shown below by an example, the initial state can also 
be chosen as the insulator limit of the disordered system, described by an 
ensemble of diagonal matrices.
Although this corresponds to a same value for all initial off-diagonal variances  
(that is, zero), however a choice of different rates of change of $h_{kl}$ 
with respect to $Y$ can result in different possible 
values for each $h_{kl}$ at a later stage.

	As an example, consider an Anderson system with a Gaussian site 
disorder (of variance $W^2/12$ and mean zero), same for each site, 
and an isotropic Gaussian hopping with a non-random component 
(of variance $W_s^2/12$ and mean $t_s$ with $s=1,2$ for real and 
imaginary parts respectively) between nearest neighbors (referred as 
ensemble {\bf G} later on).      
The corresponding probability density can be described by eq.(1) with  
\begin{eqnarray}
h_{kk} &=& W^2/12, \qquad \qquad b_{kk}=0 \nonumber \\
h_{kl;s} &=& f_1(kl;s)\; W_s^2/12, b_{kl;s}= f_2(kl;s)\; t_s  \nonumber
\end{eqnarray}
where $f_1(kl;s)=1$, $f_2(kl;s)=1$  for 
for $\{k,l\}$ pairs representing hopping,  $f_1(kl;s)\rightarrow 0$ 
and $f_2(kl;s)\rightarrow 0$ for all $\{k,l\}$ values corresponding 
to disconnected sites.  As obvious, here the distribution parameters 
$h_{kl}$ depend on more than one system parameters, 
namely, the disorder parameters $W$, $W_1$ and $W_2$ as well as 
various functions of site-indices. The latter, being invariant 
of motion, give the parameters $Y_2,..,Y_M$.  
The $Y$ for this case can be obtained by using eq.(4),  
\begin{eqnarray}
&Y&=-{N\over 2 M\gamma} \alpha + C  \\
\alpha&=&{\rm ln}|1-\gamma W^2/12| \nonumber \\
&+& (z/2)\sum_s {\rm ln}\left[|1-\gamma W_s^2/6||t_s+\delta_{t_{s}0}||
 \right] + C, 
\end{eqnarray} 
 with $M=\beta N(N+ z(1-\delta_{t0})+2-\beta)/2 
\approx \beta N^{2+\epsilon} /2$. 
Here  $z N$ is the  number of connected sites (nearest-neighbors)  which 
depends on the topology and the dimensionality $d$ of the system and the  
$\epsilon$ is a function of $z$, 
$\epsilon(z)=({\rm log}(N+z(1-\delta_{t0})+2-\beta)/{\rm log}N) -1$; 
$\epsilon \rightarrow 0$ for $z<<N$.

 Now consider an  insulator as the initial state (in the same site-basis as used for 
{\bf G}) with zero hopping, that is, $W_s=0$, $t_s=0$ and a Gaussian 
site disorder 
with variance $(W^2/12)=(2\gamma)^{-1}$ (referred as {\bf G$_0$} ). 
This corresponds to an ensemble of diagonal matrices 
with $h_{kk}=(2\gamma)^{-1}$, $h_{kl;s}=0$ for $k\not=l$  
and $b_{kl;s}=0$ for all $k,l$. A substitution of these values in 
eq.(5) gives the initial value of $Y$, say $Y_0$, where 
$Y_0=-{N\over 2\gamma M}\alpha_0+C$ with $\alpha_0=-{\rm ln} 2$. 
Note, the basis being same, the parameters $Y_j$ (for $j\ge 2$) 
are same for both {\bf G} and {\bf G$_0$}. (The advantage of choosing the above 
initial state is explained later).
As obvious, starting from {\bf G$_0$}, a variation of 
diagonal disorder $W$, hopping parameters $W_s$ and $t_s$  with  
rates 
\begin{eqnarray}
& &{\delta h_{kk}\over \delta W} = W/6, \nonumber \\
{\delta h_{kl;s}\over \delta W_s} &=& W_s f_1(kl;s)/6,  
{\delta b_{kl;s} \over \delta t_s} = t_s f_2(kl;s) \qquad k\not=l \nonumber
\end{eqnarray}
 can lead to the ensemble $G$. 
Using $(\partial Y_j/\partial x)=0$ for $j\ge 2$ 
with $x\equiv W,W_s,t_s$, and, eq.(5) to obtain 
$(\partial Y/\partial x)$, it can be seen that the above rates 
correspond to 
\begin{eqnarray}
{\partial h_{kk}\over \partial Y} &\propto& |1-\gamma W^2/12|, \nonumber \\
{\partial h_{kl}\over \partial Y} &\propto& f_1 |1-\gamma W_s^2/6|, \nonumber \\
{\partial b_{kl}\over \partial Y} &\propto& f_2 t_s;
\end{eqnarray}
 the variances and 
means of different matrix elements therefore change with different rates 
with $Y$.

The  distribution $P$ of the eigenvalues $E_n$ for a metal 
(for the energy ranges with fully 
extended eigenfunctions) is given by the Wigner-Dyson distribution, 
$P(\{E_n\}) = \prod_{i< j}| E_i - E_j|^{\beta}
{\rm e}^{-{\gamma\over 2}\sum_k E_k^2}$,
and, for an insulator by a Poisson distribution \cite{ss}. 
The  distribution for various transition stages can be 
obtained by integrating $\rho$ over the associated eigenvector space.
Let $P(\{E_n\},Y(h,b))$ be the joint probability of finding eigenvalues
$\lambda_i $  of $H$ between $E_i$ and $E_i+{\rm d}E_i$ ($i=1,2,..,N$) at
a given $h$ and $b$, it can then be expressed as
$P(\{E_n\},Y)= \int \prod_{i=1}^{N}\delta(E_i-\lambda_i) 
\rho (H,Y){\rm d}H$.
Using the above definition in eq.(3), it can be shown that the eigenvalues 
of $\rho(H)$ undergo a diffusion dynamics along with a finite drift due to 
their mutual repulsion, (see \cite{ps} also)

\begin{eqnarray}
{\partial P\over\partial Y}   =
\sum_n {\partial \over \partial E_n}
\left[ {\partial \over \partial E_n} +
\sum_{m\not=n}{\beta \over { E_m-E_n}} + \gamma E_n \right] P
\end{eqnarray}

Again the steady state of the evolution is given by the limit  
${\partial P/\partial Y}\rightarrow 0$; $P(\{E_n\})$ in this limit  
turns out to be a Wigner-Dyson distribution.

	The eq.(8) can be used to obtain the correlations between 
levels. For example, a knowledge of its solution $P$ gives  
the static correlations 
\begin{eqnarray}
R_n(E_1,E_2,..,E_n;Y)={ N! \over {(N-n)!}}\int P(\{E_j\},Y)
{\rm d}E_{n+1}..{\rm d}E_N. 
\end{eqnarray}
The $P$ can be obtained by using the analogy of eq.(8) with the equation 
governing the evolution of the eigenvalues of  Brownian ensembles (BE) of 
hermitian type \cite{dy,me}. The latter, has been studied in great detail 
in past and many of its statistical spectral properties are already 
known \cite{ap}. A brief description of the BE is given in the next section.  

	It should be noted here that the single-parametric evolution of 
the matrix elements of the AE in terms of the complexity parameter $Y-Y_0$ 
would result in a similar evolution for their eigenvector components too;
this can be shown by integrating the eq.(3) over all eigenvalues. However 
in this paper we confine ourselves to the discussion of eigenvalue statistics 
only; the details for the eigenvector statistics will be published elsewhere.

\section{Spectral Properties of Brownian Ensembles}   

	.

	The stationary random-matrix ensembles were introduced in the past to 
model quantum mechanical operators of complex systems in which a certain 
set of quantities (for example, total spin, charge or isotopic spin) was 
exactly conserved; no other integral of the motion existed even 
approximately \cite{dy,me}. The total set of the states of the system could then be 
divided into subsets, each subset corresponding to a particular set of values 
for the conserved quantities. This divides the matrix representation of the 
operator in various blocks; the deterministic uncertainty due to complicated 
nature of the interactions leads to randomization of the blocks. Due to lack 
of correlation between energy levels of states belonging to different 
subsets, different blocks are uncorrelated.  The statistics of the levels 
within one subset can then be described by a separate random matrix model 
which can be of various types based on the underlying symmetry \cite{dy,me}. 

The stationary ensembles are inappropriate models for systems possessing 
approximate conservation laws. However such systems occur more 
frequently in practice which motivated Dyson to introduce 
the Brownian ensembles (BE) of random matrices \cite{dy,me}. As the latter 
have been discussed in detail in past e.g. in \cite{me,ap} (and references therein),
here we give only a brief review of the BEs related to Hermitian matrices.   
Consider the Hamiltonian operator $H$ of a system 
with its elements given by $H_{kl}$ at "time" $\lambda$ and 
$H_{kl}+\delta H_{kl}$ at "time" $\lambda+\delta \lambda$. A Brownian 
motion of $H$ is defined by requiring that each $\delta H_{kl}$ is a random 
variable with the moments: $<\delta H_{kl;s}>=- \gamma H_{kl;s} \delta \lambda$, 
$<(\delta H_{kl;s})^2>=g_{kl} \delta \lambda$ \cite{me}. The evolution of 
the distribution of matrix elements, from any arbitrary initial state, can then 
be given by a Fokker-Planck equation which has the same form as eq.(3) with 
$Y \propto \lambda^2 $. For $\lambda \rightarrow \infty$, the distribution 
approaches steady state which corresponds to one of the stationary ensembles. 
The crossover to stationarity is rapid, discontinuous, as a function of $\lambda$, 
for infinite matrix sizes or very large energy-ranges.


	A Brownian ensemble can therefore be described as a non-stationary 
state of the matrix elements undergoing a cross-over due to a random perturbation 
of a stationary ensemble by another one. For example, in the case of Hermitian 
operators, a Brownian ensemble $H$ can be given as $H=\sqrt{f} (H_0+\lambda V)$ 
(with $f=(1-\lambda^2)^{-1}$); here $V$ is a random perturbation of strength 
$\lambda$, taken from a stationary ensemble, and applied to an initial 
stationary state $H_0$ (see also \cite{ap}). 
Using 2nd order perturbation theory, it can be shown that the eigenvalues 
$E_j$, $j=1,2,..,N$ of $H$ execute a Brownian motion too, with their evolution described 
by an equation same as eq.(8) (with $Y\propto \lambda^2 f$). 
The eigenvalue statistics (e.g. static correlations given by 
eq.(9)) of a BE can then be obtained by solving eq.(8). 
The eq.(8) is equivalent, under a Wick rotation, to the Schrodinger equation of the 
Calogero-Sutherland Hamiltonian; the equivalence has been used to obtain the eigenvalue 
correlations for many BEs \cite{ap}. It is shown moreover that the crossover in correlations 
is governed, for small $\lambda$ and large $N$, by a rescaled parameter $\Lambda$ which 
measures locally  the mean-square symmetry breaking matrix element in units of the mean 
eigenvalue spacing of $H$.  

%
%
%

 The type of a BE, appearing during the 
cross-over, depends on the nature of stationary ensembles $H_0, V$ and their different pairs 
may give rise to different BEs \cite{ap}. The present knowledge of ten types of stationary 
ensembles \cite{me} leads to possibility of many such cross-overs and, consequently, 
many types of BEs.  For example, the Hamiltonian of a disordered system or 
an autonomous chaotic system, with time-reversal symmetry, can usually be modeled 
by  Gaussian orthogonal ensemble. The breaking of time-reversal symmetry e.g. by 
switching of a magnetic field, with $\lambda$ as a measure of the breaking, 
perturbs the Hamiltonian $H_0$. The statistical behavior of the system now depends 
on the energy-range of interest. At asymptotically large energies, the 
statistics can be modeled by Gaussian unitary ensembles; however at intermediate 
energies with sufficiently small values of $\lambda$, an intermediate statistics 
(a BE between GOE and GUE) would be obtained, indicative of a non-equilibrium 
behavior. Similarly if the system is integrable with regular classical motion for 
$\lambda=0$ and fully chaotic for $\lambda\not=0$, the statistics undergoes the 
Poisson $\rightarrow$ GOE crossover; (the BE in this case is a superposition of 
Poisson and GOE ensembles). For many type of crossovers, beginning from various stationary states 
e.g. GOE, GSE, 2GOE, Poisson, uniform etc and approaching GUE in limit 
$\lambda \rightarrow \infty$, the $2^{nd}$ order correlation functions for all $\Lambda$ have 
been explicitly evaluated \cite{ap}; for the other transitions the correlations are given 
implicitly by a hierarchic set of relations \cite{ap,ps}.


Here we discuss only the BEs appearing 
during a transition from Poisson $\rightarrow$ Wigner-Dyson ensemble (referred 
as Wigner-Dyson transition or WDT) caused by a perturbation of the former by 
the latter (that is, taking $H_0$, $V$ as Poisson and Wigner-Dyson ensemble 
respectively). As this transition results in a change of localized 
eigenstates to delocalized ones,  its relevance for the study of MIT 
is intuitively suggested. The BEs related to the Poisson $\rightarrow$ 
Wigner-Dyson transition can be described by a $N\times N$ ensemble $H$ 
represented by the following probability distribution for all (independent) 
matrix elements: 
\begin{eqnarray}
\rho(H)  \propto {\rm exp}{\left[-\gamma \sum_{i=1}^N H_{ii}^2 
- 2 \gamma (1+\mu) \sum_{i<j} |H_{ij}|^2 \right]} 
\end{eqnarray}
 with $(1+\mu)=(\lambda^2 f)^{-1}$; here $H=H_0$ for $\lambda \rightarrow 0$ or 
$\mu \rightarrow \infty$. An ensemble $H$ given by the above measure,
 is also known as Rosenzweig-Porter ensemble (RPE); Note it also 
corresponds to an ensemble of Anderson 
Hamiltonians with very long range, isotropic, random hopping.  

	The eq.(8) describes the evolution of the eigenvalues of a generalized 
Gaussian ensemble with a probability density (1) and is therefore 
applicable for the BEs defined by a probability density (9) too.
A comparison of measure (9) with measure (1), gives a variance 
$h_{kl;s}= (4\gamma (1+\mu))^{-1}$, $h_{kk;s}=(2\gamma)^{-1}$ and mean 
$b_{kl;s}=0$ for all $k,l$ and $s$ indices. Using these values in eq.(4), 
the parameter $Y$ for the BE case can be given as 
 
\begin{eqnarray}
Y &=& - {1\over 2\gamma}
{(N-1)\over (N+2-\beta)} {\rm log}(1-{1\over 2(1+\mu)}) + Y_0  
\nonumber\\
&\approx& {1 \over 4 \gamma \mu} + Y_0 \qquad ({\rm for}\quad \mu >> 1)
\end{eqnarray}
with $M=\beta N (N+2-\beta)/2$ and 
 $Y_0={N\over 2\gamma M} {\rm ln}2+C$ as the complexity parameter of 
the ensemble $H_0$ 
(note, $Y=Y_0$ for $\mu \rightarrow \infty$).

	A typical matrix in the ensemble (9) has the diagonal elements 
of order $\gamma^{-1/2}$ and off-diagonals of the order of 
$(\gamma\mu)^{-1/2} (=o(Y-Y_0)^{1/2})$. 
The number of off-diagonals being $N$ times more than the diagonals, 
the matrix behavior is governed by the parameter $\mu$.  
 Thus, for large BE ($N \rightarrow \infty$),  
a radical change from Wigner-Dyson case can only occur 
if $\mu$ increases more rapidly than $N$ (which makes the total 
strength of the off-diagonals weaker than that of diagonals). 
This results in  three 
different regimes of the mean-level density $R_1(E)$ \cite{shep}: 

\begin{eqnarray} 
& & R_1(E) = {N\over \sqrt{\pi}}{\rm e}^{-E^2} \qquad  
{ {\rm for} \;\; N (Y-Y_0) \rightarrow 0} \\ 
&=& {\sqrt{8 N \gamma (Y-Y_0)-E^2} \over 4\gamma \pi (Y-Y_0)} 
\qquad {{\rm for}\; N (Y-Y_0) \rightarrow \infty} \\ 
&=& N F(E,a) \qquad {{\rm for}\quad  N (Y-Y_0)=a}  
\end{eqnarray}

with $a$ as an $N$-independent constant. Although the exact form 
of the function $F(E,a)$ is not known, its limiting behavior can 
be given as follows: 
$F(E,a) \approx {\rm e}^{-E^2}/\sqrt{\pi}$ for $a << 1$ and 
$F(E,a) \approx (4\pi\gamma a)^{-1}\sqrt{8\gamma a-E^2}$ for 
$a>>1$, $E^2 << a$ \cite{shep}.  

	The $1^{\rm st}$ order correlation $R_1$, also known as mean level density, 
changes from  an exponential to 
semi-circular form at the scale of $(Y-Y_0)\sim N\Delta_l^2$ with $\Delta_l$ as the 
local mean level spacing; the evolution of $R_1$ can therefore be described in terms of the 
parameter $(Y-Y_0)$. However the transition of higher order correlations $R_n$ $(n>1)$ 
occurs at a scales determined by $(Y-Y_0) \sim \Delta_l^2$ \cite{ap,shep,ps}.  
As a result, their transition to equilibrium, with $|Y-Y_0|$ as the evolution 
parameter, is rapid, discontinuous for infinite dimensions of matrices \cite{dy}. 
But for small-$Y$ and large $N$, a smooth crossover can be seen in 
terms of a rescaled parameter $\Lambda(E)$: 
\begin{eqnarray}
\Lambda(E,Y)= |Y-Y_0|/\Delta_l^2
\end{eqnarray}
For finite $N$, $\Lambda$ varies smoothly with changing $\mu$: $\Lambda=R_1^2/4\gamma \mu$.   
This results in a continuous family of BEs, parameterized by $\Lambda$, 
existing between Poisson and Wigner-Dyson limit. 
However the level-statistics for the large
BE ($N \rightarrow \infty$) can be divided into three regions \cite{shep}:  

(i) {\bf Poisson regime: {$N^2(Y-Y_0) \rightarrow 0$}}: 
The off-diagonals, responsible for the correlation between levels, are 
negligible. The lack of repulsion between levels results in a mean level spacing 
 $\Delta_l \propto N^{-1}$ (see eq.(12)), thereby, giving   
$\Lambda \rightarrow 0$ and the Poisson statistics.

\vspace{0.1in}

(ii) {\bf WD regime: $N^2(Y-Y_0) \rightarrow \infty$}: 
The contribution from both, the diagonals as well as off-diagonals is of 
the same order, leading to long-range correlations between levels. 
The repulsion of levels now results in a mean level spacing $\Delta_l \propto N^{-1/2}$. 
(see eq.(13)) which gives $\Lambda \rightarrow \infty$ and  Wigner-Dyson 
statistics.  

\vspace{0.1in}

(iii){\bf Critical regime: $N^2(Y-Y_0)=(4\gamma c)^{-1}$= a constant}: 
For $\mu=c N^2$  with $c$ as a constant independent of $N$, a sequence 
of approximately $o(1/\sqrt{c})$ levels show Wigner-Dyson behavior. The 
more distant levels display weak correlations of the type existing near 
the Poisson limit resulting in a $\Delta_l \approx o(1/N)$. 
The parameter $\Lambda$ is therefore $N$-independent: 
\begin{eqnarray}
\Lambda(E) = (1/4 c\pi\gamma){\rm e}^{-E^2}; 
\end{eqnarray}
note it is also independent of the 
symmetry parameter $\beta$. 

The finite, non-zero $\Lambda$-value for $\mu=c N^2$ in limit 
$N\rightarrow \infty$ therefore gives rise to a third statistics, 
intermediate between Poisson and Wigner-Dyson ensemble, which is known as 
the critical Brownian ensemble (CBE). This being the case for arbitrary 
values of $c$ (non-zero and finite), an infinite family of critical BE, 
characterized by $c$ (or $\mu_c=c N^2$), occurs during WDT. 
Note that the critical BEs, with $c\rightarrow \infty$ and 
$c\rightarrow 0$, correspond to the Poisson and Wigner-Dyson limit, respectively.

 The presence of a family of the critical BEs can be seen from any of the fluctuation 
measures for WDT. One traditionally used measure 
in this regard is the relative behavior of the tail of nearest-neighbor 
spacing distribution $P(s,\Lambda)$, defined as 
$\alpha (\delta,\Lambda) = \int_0^\delta (P(s,\Lambda)-P_w(s))
{\rm d}s/\int_0^\delta (P_p(s)-P_w(s)){\rm d}s$ with $\delta$ as any 
one of the crossing points of $P_w(s)$ and $P_p(s)$ 
(here subscript $w$ and $p$ refer to the Wigner-Dyson case and Poisson 
case respectively) \cite{hs}.  
In the limit $N\rightarrow \infty$, $\alpha=0$ and $1$ for Wigner-Dyson 
and Poisson limit respectively.  
The figure 1 shows the numerically obtained behavior of $\alpha$ 
 (for $\delta\approx 2.02$)  with respect 
to $|z-c|$ (=$|\mu-\mu_c|N^{-2}$) for a fixed $c$ (arbitrarily chosen) 
with $z$ as a variable; Here $z$ and $c$ are the values of the parameter 
$\mu N^{-2}$ for a general BE and a critical BE respectively. 
The constant value of $\alpha$ at $|z-c|=0$ for different 
$N$-values confirms the size-independence of the level-statistics of 
BE with parameter $\mu=c N^2$ and therefore its critical nature; 
we have  verified it for other $c$-values too and find, for finite, non-zero $c$-values, 
$0<\alpha<1$. Further the 
convergence of $\alpha$-values for BEs with different 
$\mu$ and $N$-values on two branches indicates the presence of a scaling 
behavior in the level-statistics of BEs with $|z-c|$ (=$|\mu-\mu_c|N^{-2}$)
as the scaling parameter.
	
	As shown in figure 1(a), $\alpha$ for a critical BE is between $0$ and $1$. 
A fractional value of $\alpha$ indicates a tail-behavior of critical  
BE different from that of Poisson as well as Wigner-Dyson limit. 
As  shown in figure 1(b), the $P(s)$ for a critical BE with a finite parameter 
$c$ has an exponential tail,  $P({\rm large}\; s) \sim \exp(-\kappa s)$; 
this behavior of $P(s)$ is also sometimes referred as semi-Poisson distribution, due to 
presence of repulsion at small energy-scales and exponential decay at large separations.

\section{Analogy between Brownian Ensembles and Anderson Ensembles}

The same evolution equations of $P$ for AE and BE imply a similarity  
in their eigenvalue distributions for all $Y$-values,  
under similar  initial conditions (that is, $P(\mu,Y_0)$ same for 
both the cases). As a result, one obtains the analogous evolution equations for 
their correlations $R_n$ too.  The mean level density $R_1(E,Y-Y_0)$ of 
an AE can therefore be given by the level-density of a BE with same 
$|Y-Y_0|$ value (and appearing during a Poisson to Wigner-Dyson transition).  
Similarly, the analogy of evolutions of $R_n$ ($n>1$) in the two cases implies  

(i) a smooth crossover of $R_n$ for finite size Anderson systems in arbitrary dimensions,  

(ii) the parameter $\Lambda$ governing the smooth crossover of $R_n$ for finite size AEs 
can again be defined by eq.(15), with $Y-Y_0$ given by eq.(4) and $\Delta_{l}$ 
as the local mean level spacing for AE (see \cite{pp} also),

(iii) the correlations $R_n$, $n>1$, of an AE can 
be given by those for a BE with a same $\Lambda$ value 
although their parameters $Y$ (as well as level densities)  may be different; 

(iv) the discontinuity of the transition of $R_n$ for infinite size of Anderson matrix, 

(v) the existence of a size-invariant level statistics, different from two end-points, 
if an AE has $\Lambda={\rm size-independent}$; the statistics 
 survives the thermodynamic limit 
$L\rightarrow \infty$. As explained later by an example, the above condition on $\Lambda$ 
is satisfied at the critical point of $d>2$ dimensional Anderson Systems; the corresponding 
level-statistics is referred as critical.
.
                                                                                                     
	The implications (i,iv)  are well in agreement with known results about 
AE-correlations \cite{kkk}. The implications (ii,iii,v) indicate the single parametric 
dependence of the level statistics for AEs. 
The parameter $\Lambda$ for the AE and BE will henceforth be referred as 
$\Lambda_a$ and $\Lambda_b$, respectively.
The level-statistics of a finite-size AE at $\Lambda_a$ is then 
given by a BE with its parameter $\mu$ satisfying the condition 
\begin{eqnarray}
\Lambda_a=\Lambda_b
\end{eqnarray}
 where $\Lambda_b={R_1^2/4\gamma\mu}$ with $R_1$ 
as the level-density of the BE; the determination of $\Lambda_a$ 
is explained later by using an example. 
As the BEs with different combinations of the parameters 
$\mu$ and $N$ can have same $\Lambda_b$, the correlations of a 
finite size AE can be mapped to many BEs. However 
the critical BE corresponding to a critical AE is unique; this can 
be understood as follows. 
Using eq.(16) in eq.(17), the parameter $c$ for a critical BE corresponding 
to an AE can be given by 
\begin{eqnarray}
c=(4\pi \gamma \Lambda_a)^{-1} {\rm e}^{-E^2}.
\end{eqnarray}
The $\Lambda_a$ for a critical AE being size-independent, 
its critical BE analog remains same for all system sizes. 
However, the  $\Lambda_a$ for an AE, away from 
its critical point, is size-dependent and therefore corresponds to different 
$c$ values (that is, different critical BEs) for different system sizes. 

The $\Lambda_a$ for a disordered system can be determined by a knowledge of $Y-Y_0$ 
and $\Delta_l$.  The complexity parameter $Y-Y_0$ is system-specific and 
depends on various system parameters.  For a $d$-dimensional disordered 
system of linear size $L$, the local mean level spacing $\Delta_l$ 
within a correlation volume of linear dimension $\zeta$ is  
related to mean level density $R_1$: 
$\Delta_l=(L/\zeta)^d R_1^{-1}$ where 
$\zeta$ is the localization length or correlation length in case of 
localized states and extended states, respectively \cite{pp1,kkk}. 
The $\zeta$ can be determined by a knowledge 
of the wave-function correlations e.g. inverse participation ratio $I_2$ 
\cite{fyd}: $\zeta^d \propto (I_2)^{-1}$ for localized eigenstates \cite{fyd}. 
As mentioned in the last paragraph of section I, the wave-function statistics 
and, therefore $\zeta$, can also be described, in principle, by a complexity parameter 
formulation. However, the related work being still in progress, we use, in this paper, 
the $\zeta$ results given by previous studies. 

Let us consider the example {\bf G} given in section 2; its parameter 
$Y$  is given by eq.(5). The initial state {\bf G$_0$} has the parameter $Y_0$ same as 
that of the initial state chosen in the BE case in Section II.  
Note as $N |Y-Y_0| \rightarrow$ an $N$-independent function for the case {\bf G}, 
its $R_1$ is given by eq.(14) with $ R_1= N F$. Using eq.(5) 
in eq.(15),  $\Lambda$ for the case {\bf G} can be given as 
\begin{eqnarray}
\Lambda_a(E,Y) &=& \left(|\alpha-\alpha_0| F^2\over \beta \gamma\right)
\zeta^{2d} L^{-d} 
\end{eqnarray}  
with $F(E)$ giving the energy-dependence of $\Lambda$ (as $|\epsilon|\approx 0$ 
for large $N$). Following eq.(17), the level-statistics at $\Lambda_a \rightarrow 0, \infty$ 
corresponds to Poisson (or insulator limit) and Wigner-Dyson behavior (metallic limit), 
respectively. In finite systems, a change of disorder results in a smooth 
variation of $\zeta$ as well as $\alpha-\alpha_0$ and, therefore, $\Lambda_a$  
which induces a crossover of the level-statistics from 
Poisson $\rightarrow$ Wigner-Dyson ensemble. The intermediate 
level-statistics at each $\Lambda_a$ of a finite-size AE is then 
given by a $N_1\times N_1$ BE with its parameter $\mu$ satisfying 
the relation $\Lambda_a=\Lambda_b$:

\begin{eqnarray}
\mu \approx  \beta (4 \pi |\alpha-\alpha_0| F^2)^{-1} \zeta^{-2d}
L^{d} R_1^2 
\end{eqnarray}
with $R_1\equiv R_1(E;\mu,N_1)$ as the level-density of the BE. As obvious, 
the determination of $\mu$ from the above equation is not easy, its both 
sides being $\mu$-dependent. The $R_1$ for the critical BEs being $\mu$-independent 
(given by eq.(12)), it is preferable to map an AE to a critical BE; 
the substitution of eq.(19) in eq.(18) gives the $c$-parameter for the 
corresponding critical BE:

\begin{eqnarray}
c \approx  \beta (4 \pi |\alpha-\alpha_0| F^2 {\rm e}^{E^2})^{-1} \zeta^{-2d}  
L^{d} 
\end{eqnarray}
Thus  each state of disorder in an AE of size $L$ ($N=L^d$) can be 
mapped to a critical BE with the parameter $c$ given by eq.(21). 
Note, the right side of eq.(21) being energy-dependent, different 
energy ranges of a given AE will, in general, correspond to different 
critical BEs.

	Equation (19) indicates the sensitivity of the parameter $\Lambda_a$ to 
localization length $\zeta$ and  system size $L$. 
It is now well-known that  $\zeta$ is a function of disorder-strength, energy, 
system-size $L$ as well as the dimensionality of the system. 
For systems with finite $L$ (in arbitrary dimensions $d\ge 1$), $\zeta$, at a fixed energy, 
decreases with increasing disorder-strength. Consequently, in the strong disorder limit 
(where $\zeta \sim o(L^0)$), $\Lambda_a \rightarrow 0$ and  the level-statistics of the AE 
approaches Poisson behavior (as $\Lambda_b \rightarrow 0$ for its BE analog). 
In the opposite limit $\zeta \sim o(L)$ of weak disorder,  
$\Lambda_a\rightarrow \infty$ and, therefore,  
the statistics of the AE is given by a BE at $\Lambda_b \rightarrow\infty$ 
which corresponds to Wigner-Dyson behavior. By a suitable choice of disorder, however,  
it is possible to achieve  finite values of the ratio $\zeta^2/L$ (due to finite $L$) 
in arbitrary 
dimensions which in turn gives finite, non-zero $\Lambda_a$ and, thus, a finite $c$ 
for its BE analog. The latter implies that the AE-statistics  
 is intermediate between Poisson and Wigner-Dyson 
limit, with an exponential decay of the tail of its nearest-neighbor spacing distribution 
$P(s)$. For finite $L$, therefore, a smooth crossover from Poisson to Wigner-Dyson
statistics can be seen, {\it for any dimensionality} $d \ge 1$, as a function of 
$\Lambda_a$ by varying the disorder-strength. 
Note, two finite size AEs of different dimensions can show same level-statistics if their 
parameters $\Lambda_a$ are equal. For example, consider the behavior of levels of a 
one-dimensional AE of size $L$ and  at a disorder strength which gives $Y=Y_1$. 
The behavior will be same as that of a 
three-dimensional AE of linear size $L$, at a disorder strength  which gives $Y=Y_3$ where
$Y_3=Y_1 (\zeta_1 R_1^{(1)}/\zeta_3 R_1^{(3)})^2$; here $R_1^{(d)}$ and $\zeta_d$ refer to  
the mean level density and localization length in dimension $d$, respectively.

	The dimensionality dependence of the Anderson transition and the critical 
level-statistics is well known. For example, the  level statistics at the critical 
disorder for $d>2$ dimensional, finite systems shows a "semi-Poisson" 
behavior which survives the infinite size limit. The same behavior is seen for 
$d \le 2$ dimensional finite systems, in a regime where $\zeta \sim L$, however 
the statistics approaches a Poisson behavior in the thermodynamic limit.   
The above behavior can be explained within "$\Lambda$-formulation".
As mentioned in the section II, a "semi-Poisson" behavior of the level-statistics is a 
characteristic of critical BEs with finite $c$ parameters and therefore of the AEs with 
finite $\Lambda_a$ parameters (see eq.(19)). The AE-statistics  is expected 
to maintain its semi-Poisson behavior even in thermodynamic limit if 
$\Lambda_a$=size-independent. In this sense, $\Lambda_a$ can be identified with 
the dimensionless conductance $g$: 
both $g, \Lambda_a \rightarrow 0, \infty, {\rm constant}$ correspond to same statistical 
limit, namely, Poisson, Wigner-Dyson and a critical level statistics, respectively.
In fact, $\Lambda_a$ can be expressed in terms of the 
dimensionless conductance $g$ of the system. This is because $g$ is connected 
to $\zeta$ (based on scaling theory of localization for disordered systems \cite{lr}): 
$\zeta \propto L {\rm log} g^{-1}$ for exponentially localized states, 
$\zeta \propto |(g/g_c)-1|^{-\nu}$  near the critical point and 
$\zeta \sim o(L)$ with $g \propto L^{d-2}$. For example, using eq.(19), 
the $\Lambda_a-g$ relation for the case $G$ near the critical point can be given as: 
\begin{eqnarray}
\Lambda_a(E,Y) = \left(|\alpha-\alpha_0| F^2 \zeta_0^{2d}\over \beta \gamma\right)
|(g/g_c)-1|^{-2\nu d} L^{-d} 
\end{eqnarray}  
with $g_c$ as the critical point conductance and $\nu$ as the critical exponent.

	As indicated by eq.(19), the size-independence of $\Lambda_a$ is governed 
by the size-dependence of the localization length.  
For example, for a d-dimensional disordered system, 
with  $|Y-Y_0|\simeq o(L^{x_1 d})$, the level-density is $R_1(Y-Y_0)\simeq o(L^{x_2 d})$ 
where 
\begin{eqnarray}
x_2 &=& 1   \qquad \qquad  {\rm for}\qquad x_1 \le -1, \nonumber \\ 
x_2 &=& (1-x_1)/2  \qquad\qquad {\rm for} \qquad x_1 \ge -1. 
\end{eqnarray}
Thus the critical point of the level-statistics (that is, $\Lambda$=size-independent) 
can exist only if, in thermodynamic limit, 
the disorder conditions in the system give rise to a localization 
length $\zeta \sim o(L^{x_3})$ where  $x_3 \approx (2-x_1-2 x_2)/2$
or, equivalently, 
\begin{eqnarray}
x_3 &=& |x_1|/2  \qquad \qquad  {\rm for} \qquad x_1 \le -1 \nonumber \\
x_3 &=& 1/2  \qquad \qquad {\rm for}  \qquad x_1>-1.  
\end{eqnarray}
The existence or non-existence of a critical level statistics in an AE 
therefore depends on the the size dependence of $\zeta$ which in turn is sensitive to 
the dimensionality  of the system\cite{lr}: 

{\bf {Case $d \le 2$}}: For a $d=1$ disordered lattice,  
almost all states are known to be exponentially localized even in a weak disorder 
limit. The  $\zeta$ in this case is finite, 
$\zeta \approx \pi l \sim o(L^0)$ \cite{lr} (with $l$ as the mean free path), 
which gives $x_3=0$. As obvious, the condition (21) can not be satisfied for 
any $x_1$, equivalently, for any $Y-Y_0$ (e.g. for any disorder 
conditions). As a consequence, a critical level-statistics can not occur in 
one-dimensional case.

  Equation (16)  suggests that a "semi-Poisson" type statistics can be seen
in $d=1$ case for  $L$ of the order of few mean free paths (i.e for finite $l^2/L$). 
In limit $L\rightarrow \infty$, however,  $\Lambda_a \rightarrow 0$ and 
the level statistics  approaches Poisson behavior 
irrespective of the disorder strength. 

 In two dimensions, the perturbative estimate of the localization length is 
$\zeta \approx l {\rm exp}[\pi k_F l/2]$ with $k_F$ as the Fermi wave number \cite{lr} 
and,  in the limit $L\rightarrow \infty$, electronic states are expected to be localized 
even for small microscopic disorder \cite{mk}. This again corresponds to $x_3=0$ (thus 
absence of critical level-statistics), and, the Poisson statistics for the levels  
in thermodynamic limit (as $\Lambda_a \rightarrow 0)$. Note, however, that due to exponential 
nature of $\zeta$, the ratio $\zeta^2/L$ can be kept non-zero and finite (by changing disorder) 
for a large range of $L$. The system can therefore show the semi-Poisson statistics in a 
large range of system sizes.

{\bf {Case $d>2$}}: For $d>2$ dimensional, infinite systems, the change 
of disorder  $W$  leads to a discontinuous 
change in $\zeta$ and thereby $\Lambda_a$: $\zeta \propto |1-(W/W_c)|^{-\nu}$. Here 
$\nu$ is the critical exponent and $W_c$ is the critical disorder. For $W >W_c$, almost all 
states are exponentially localized  with $\zeta \sim o(L^0)$ which results in 
$\Lambda_a \rightarrow 0$ and Poisson behavior of the statistics. 
For $W < W_c$, the delocalization of states occurs with $\zeta \rightarrow o(L)$; 
this gives $\Lambda_a \rightarrow \infty$ and the Wigner-Dyson statistics.  
At $W_c$, however, the inverse participation ratio $I_2$ for $d>2$ case 
shows  an anomalous scaling with  $L$ \cite{weg}: 
$I_2 \propto L^{-D_2}$ with $D_2$ as the multifractality exponent. This gives 
$\zeta^d \propto <I_2>^{-1} =\zeta_0^d L^{{\tilde D_2}}$ or 
$x_3={\tilde D_2}/d$ with 
$\zeta_0$ as a size-independent function; note ${\tilde D_2}=D_2$ at the critical point 
\cite{em,ver}. The size-independence of the level statistics at 
the critical disorder therefore requires  
\begin{eqnarray}
D_2 &=& d|x_1|/2  \qquad {\rm for}\qquad x_1 \le -1 \nonumber \\
&=& d/2 \qquad {\rm for} \qquad x_1 >-1.  
\end{eqnarray}
For example, as $x_1 \approx -1$ in the case {\bf G},  the existence  
of its critical point requires $D_2 \approx d/2$. 
Note the numerical results for 
$D_2$, at the critical point of a $d=3$ dimensional AE system (of type ${\bf G}$)
fluctuate in the range 1.4-1.6 \cite{bran,ckl,ro1,ter} 
(also see references [77,79] in \cite{ro1}); this is in close agreement 
with the result given by eq.(25) for $d=3$ case. 
The above prediction for $D_2$ can be used to determine the   
critical BE analog for the critical state of the AE example $G$ for $d>2$ 
case:   
\begin{eqnarray}
c \approx  (4 \pi |\alpha-\alpha_0|\zeta_0^{2d} F^2 {\rm e}^{E^2})^{-1} \beta   
\end{eqnarray}
(as $\epsilon \approx 0$).
Thus, unlike the $d=1$ case showing only Poisson level-statistics in the 
thermodynamic limit, 
the energy levels of an infinite size AE for $d>2$ case can show three types of  
behavior, namely, Poisson, Wigner-Dyson and a  critical BE type statistics, at the 
disorder strength above, below and at the critical disorder, respectively. 

	The study \cite{ckl} suggest a connection between $D_2$ and the 
level-compressibility $\chi$: 
\begin{eqnarray}
D_2=d(1-2\chi)
\end{eqnarray}
A comparison of eq.(25) with eq.(27) 
gives the $\chi$ for $d>2$ dimensional AE at the critical point: 
$\chi \approx 0.25$. 
(Note the above $\chi$-result is valid only for the cases of type ${\bf G}$
with $\zeta\propto L^{D_2/d}$).        
The tail of the distribution $P(s)$ is also 
believed to be related to $D_2$ \cite{azks}: $P({\rm large}\;s) \approx 
{\rm e}^{-\kappa s}$ where 
$\kappa= (2\chi)^{-1} \approx 2$.  The results for $\chi, \kappa$ are in 
close agreement with earlier numerical studies on Anderson systems 
\cite{bran,ckl,ro1,ter,hs1}; our numerical study, given in section V, also confirms the 
above results. The symmetry-independence of our theoretical prediction for $\chi$ and $\kappa$ 
for Anderson systems is also in agreement with numerical observations \cite{hs1,ter}.   
As discussed later, however, the eq.(27) (and therefore above $\chi,\kappa$ results) 
seems to be valid only in the weak multifractality limit i.e $D_2 \sim d$ 
(see paragraph below eq.(36)).

For disordered systems, in general, both $Y-Y_0$ as well as  $\zeta$ are a function of 
coordination number,  disorder strength, hopping rate, dimensionality as well as 
boundary conditions of the lattice. The changing complexity due to change of the 
system parameters plays the role of a random perturbation, of strength $\sqrt{Y-Y_0}$, 
applied to the system.
Here, again, the statistics of the levels is governed by $\Lambda$ and, therefore, 
by the competition between local mean-level spacing 
$\Delta_l$ and the perturbation strength $Y-Y_0$. 
 The perturbation  mixes fewer levels with 
increasing system size if $\Delta_l$ increases with $L$ at a rate faster then 
that of $\sqrt{Y-Y_0}$ and, as a consequence, leaves the 
 level-statistics unperturbed in limit $L\rightarrow \infty$. In the opposite case with 
slower rate of change of $\Delta_l$ with $L$ (as compared to $\sqrt {Y-Y_0}$), 
even a small change in the complexity parameter is capable of mixing the levels 
 in an increasingly large energy range of many local mean level-spacings. 
This results in an increasing degree of the eigenfunctions delocalization
 and Wigner-Dyson behavior of level-statistics in thermodynamic limit.
The critical regime occurs when  both $\sqrt{Y-Y_0}$ and $\Delta_l$ change at a same 
rate with $L$; the perturbation in this case mixes only a finite (non-zero), fixed 
number of levels even when the system is growing in size. 
As $\Lambda$ remains finite in limit
$L\rightarrow \infty$, it gives rise to a new statistics different from the two 
end-points ($\Lambda \rightarrow 0$ and $\infty$). 
Note,  the disordered systems with different  dimensionality  
can have different critical values of $\Lambda$ (due to dimensionality 
dependence of $\Delta_l$ as well as $|Y-Y_0|$) and, 
therefore, correspond to critical BE analogs with different $c$ values. 
Further the boundary conditions/ topologies, 
leading to different sparsity and 
coordination numbers, can also result in different critical level statistics  
even if the underlying symmetry and the dimensionality is same; this is
in agreement with numerical observations \cite{bran} and analytical study for
2D systems \cite{ky}. A knowledge of $\Lambda$ can then be used to map 
the critical level statistics at MIT for various 
dimensions $d >2 \rightarrow \infty$ to the infinite 
family of critical BEs.

\section{Determination of fluctuation measures for MIT}:

Many results for the spectral fluctuations of the WDT with Poisson 
ensemble as an initial state are already known \cite{ap} and can 
directly be used for the corresponding measures for the MIT in 
different disordered systems.  

\subsection{MIT With No Time-Reversal Symmetry}

The fluctuation measures, for the Anderson transition in presence of a 
magnetic field, can be given by the BEs appearing during a WDT which 
violates time-reversal symmetry. Such a WDT, occurring in a complex 
Hermitian matrix space (that is $\beta=2$), corresponds to a transition 
from Poisson $\rightarrow$ GUE ensembles.  

\vspace{.1in}  

{\bf (i)  The 2-Level Density Correlator $R_2(r;\Lambda)$}:
\vspace{.1in}  
The $R_2$ for BEs during Poisson $\rightarrow$ GUE transition 
has been obtained by various studies \cite{ap,ks,fgm}. Here we 
use the form given in \cite{ks} for the purposes to be explained later   
(note, our $\Lambda$ is equivalent to $\Lambda^2/2$ used in the \cite{ks}),
 
\begin{eqnarray}
R_2(r;\Lambda)=1 + {4 \Lambda\over r} 
\int_0^\infty {\rm d}u \; F \; {\rm e}^{-2\Lambda u^2-4\pi\Lambda u}  
\end{eqnarray} 
with 
\begin{eqnarray}
F &=&   {\rm sin}(ur) f_1 - {\rm cos}(ur) f_2 \nonumber \\  
f_1 &=& (2/z)[I_1(z)-\sqrt{8u/\pi}I_2(z)] \nonumber \\
f_2 &=& (1/u)[I_2(z)-\sqrt{2u/\pi}I_3(z)]
\end{eqnarray}  
where $z=\sqrt{32\pi \Lambda^2 u^3}$ and $I_n$ as the $n^{\rm th}$
Bessel function. (Note, the eq.(4.15) in \cite{ks} has a  
misprint in the coefficient of $u$ is the exponent; the 
correct coefficient is given in the eq.(28) above ).

The eq.(28) gives the exact form of two-point correlation for 
the Anderson transition with no time-reversal symmetry.   
Here $R_2(r,\infty)= 1- {({\rm sin}^2(\pi r)/\pi^2 r^2)}$
 and $R_2(r,0)=1$ corresponding to metal and insulator regime
respectively. A substitution of critical value of $\Lambda_a$ 
in eq.(28) will thus give $R_2$  for the critical AE.  

For large $\Lambda$-values (for all $r$), $R_2$ 
can be approximated as follows \cite{ks,fgm}: $R_2=1-Y_2$ where 

\begin{eqnarray}
Y_2(r,\Lambda) &=& {\frac {- 4\Lambda}{16\pi^2\Lambda^2+r^2}} -
{\frac {1}{2\pi^2 r^2}} 
[{\rm cos}(2\pi r) {\rm e}^{-\frac {r^2}{2\Lambda}} - 1]\\
&\approx& { \frac{3}{2\pi^2 \Lambda}}  
   \frac{{\rm sin}^2(\pi r)}{{\rm sinh}^2(r\sqrt{3/2\Lambda})}
\qquad ({\rm for}\quad r << \sqrt\Lambda)         \nonumber 
\end{eqnarray} 
However, for $ r > {\sqrt\Lambda}$, 
$Y_2 (r,\Lambda) = -{\frac {4\Lambda}{16\pi^2\Lambda^2+r^2}} +
{\frac {1}{2\pi^2 r^2}}$. As  
 $\Lambda=(4c\pi\gamma)^{-1}$ (near $E=0$) for a critical BE, 
the $Y_2 \approx {(1-8\pi^2\Lambda)/2 \pi^2 r^2}$ for 
$r>2 \beta \pi\Lambda$ (here $\beta=2$).

	The above large $r$-behavior of $Y_2(r;\Lambda)$ at 
$\Lambda=\Lambda_b$ results in a non-zero, fractional value of the sum  
I=$\int_{-\infty}^{\infty} Y_2 (r;\Lambda) {\rm d}r$ for a critical BE of 
complex-Hermitian type:
\begin{eqnarray}
I\approx 1-(\beta \pi^2 \Lambda)^{-1}.
\end{eqnarray}  
Note a $0<I<1$ value is believed to be an indicator of the multifractality of 
the wavefunctions and the fractional compressibility of the spectrum;
($I=1,0$ for the WD and Poisson case, respectively) \cite{ckl,km}. 
A fractional behavior of $I$ and the multifractality is already known to 
exist in critical AE \cite{ckl,km}. Using $\Lambda_a=\Lambda_b$ 
in eq.(31), one can now determine the 
measure $I$ for an AE: $I\approx 1-(\beta \pi^2 \Lambda_a)^{-1}$.

\vspace{.1in}  
{\bf (ii) Nearest Neighbor Spacing Distribution P(S)}

\vspace{.1in}  
The nearest-neighbor spacing distribution $P(s)$ for the MIT 
with no time-reversal symmetry can similarly be given by using the one for the BE  
during Poisson $\rightarrow$ GUE transition \cite{to,lh}:
\begin{eqnarray}
P(s;\Lambda) \propto {s\over \sqrt{2\pi\Lambda}}  {\rm e}^{-s^2/8\Lambda}
\int_0^\infty {\rm d}x\; {\rm e}^{-x-x^2/8\Lambda} 
{{\rm sinh}(xs/4\Lambda)\over x}. 
\end{eqnarray}
A substitution of $\Lambda \rightarrow \infty$ and $\Lambda \rightarrow 0$ 
in the above equation gives the correct asymptotic limits, namely, Wigner-Dyson and 
Poisson, respectively: 
$P(s;\infty)=P_w(s) =32 s^2 {\rm e}^{- 4 s^2/\pi}/\pi^2$ (WD limit) and 
$P(s;0)=P_p(s)\propto {\rm e}^{-s}$ (Poisson limit). 
(Although this result is rigorous for $2\times 2$ matrix space but is 
proved reliable for systems with many levels; see \cite{lh}).

\vspace{.1in}  
{\bf (iii) Level-Compressibility}

\vspace{.1in}  
The level compressibility $\chi= 1- \int_{\infty}^{\infty} Y_2(r) {\rm d}r = 
1-I$ is an important characteristic
of the critical level statistics and the multifractal nature of the 
wavefunctions.

The $\chi$ for a BE can be obtained by using eq.(28),
\begin{eqnarray}
\chi(\Lambda) &=& 1-4 \pi \Lambda\int_0^\infty {\rm d}u
f_1 (z){\rm exp}[- 2\Lambda u^2 - 4\pi\Lambda u] \\
&\approx& 1-4\pi^2 \Lambda  \qquad {\rm for \;small} \; \Lambda \\
&\approx& (2\pi^2 \Lambda)^{-1} \qquad {\rm for \;large} \; \Lambda
\end{eqnarray}
The substitution of eq.(16) for $\Lambda_b$ in eqs.(29,30) gives, in the band around $E=0$, 
 $\chi=1-(\pi/\gamma c)$ 
and $\chi \approx (4\gamma c/2 \pi)$, in small and large $\Lambda$ limits, respectively.  
Thus  a critical BE characterized by a finite $c$ value 
shows a fractional level-compressibility.  
As clear from the above, $\chi \rightarrow 1$ for $\Lambda \rightarrow 0$ 
(or $c\rightarrow \infty$)  which corresponds  to a Poisson behavior, and,
$\chi \rightarrow 0$ for $c \rightarrow 0$ or $\Lambda \rightarrow \infty$ 
which corresponds to the GUE statistics. 

The compressibility of the energy levels of Anderson systems at their critical point 
is already known to be fractional, with $\chi=0, 1$ in the metallic  and the 
insulator phase, respectively. 
The existence of a fractional $\chi$ for both critical BE and critical AE is 
consistent with our claim about their spectral analogy. The compressibility of 
the AE with different types of disorders and lattices can now be obtained 
just by finding the same for their critical BE analogs.

	For the critical BE case ($d=1$) with large $\Lambda_b$ (equivalently, small $c$), 
eq.(27) along with eq.(35) gives 
\begin{eqnarray}
D_2 &=& 1-4\gamma c/ \pi  \qquad {\rm for}\; {\rm small}\; c 
\end{eqnarray}
The eq.(36) gives the correct fractal 
dimension in the limit $c \rightarrow 0$ (the Wigner-Dyson limit): $D_2(c=0)=1$. 
However, for small $\Lambda_b$ (or large $c$ ), eq.(27) implies  $D_2 = 2\pi/\gamma c -1$  
and therefore $D_2=-1$ in the Poisson limit $c\rightarrow \infty$,  
which is different from the expected result $D_2=0$ for the localized states. 
As mentioned in \cite{em}, similar violation of eq.(27) is indicated by numerical data 
for the tight-binding models in dimensions $d>4$. 
The observed inaccuracy of eq.(27), for both AE as well as BE
in the strong multifractality limit, also lends credence to our claim regarding 
AE-BE analogy.      

	It is worth mentioning here that, similar to the AE-BE mapping, the spectral 
statistics of any 
generalized Gaussian ensemble, with probability density given by (1), can be 
mapped to BEs \cite{ps}. The non-validity of eq.(27) for critical BEs with 
small $c$ parameters implies, therefore, the same violation for all generalized 
Gaussian ensembles in strong multifractality limit. The implication is already 
known to be correct for the  power law random banded matrices \cite{em,ver,cu1} 
(also see section VI) and for the random matrix ensemble  introduced by 
Moshe, Neuberger and Shapiro (later referred as MNS model)\cite{mns}. 

\subsection{MIT With Time-Reversal Symmetry}

 The statistical measures for the Anderson transition in
presence of a time-reversal symmetry can similarly be obtained by using their
 equivalence  to a WDT preserving the same symmetry, that is, a transition 
from Poisson $\rightarrow$ GOE ensembles; the later occurs in a real-symmetric matrix 
space (here $\beta=1$). However due to the technical difficulties \cite{ap}), 
only some approximate results are known for the latter case.

\vspace{.1in}  
{\bf (i) The 2-Level Density Correlator $R_2(r;\Lambda)$}:
\vspace{.1in}  

 The $R_2$ for small-$r$
can be obtained by solving eq.(17) of \cite{ps} for $\beta=1$ which gives 
\begin{eqnarray}
R_2(r,\Lambda) \approx (\pi/8\Lambda)^{1/2} r {\rm e}^{-r^2/16 \Lambda}
I_0(r^2/16\Lambda)
\end{eqnarray}
with $I_0$ as the Bessel function.  

Similarly for large-$r$ behavior,
$R_2$ can be shown to satisfy the relation (see eq.(23) of \cite{ap})

\begin{eqnarray} 
R_2(r,\Lambda) &\approx& R_2(r,\infty) + \nonumber \\
&+& 2\beta \Lambda
\int_{-\infty}^{\infty} {\rm d}s {{R_2(r-s;0)-R_2(r-s;\infty)\over
(s^2 + 4\pi^2 \beta^2 \Lambda^2 )}}\\
&\approx& R_2(r,\infty) +
{2\beta \Lambda /( r^2 + 4\pi^2 \beta^2 \Lambda^2)}
\end{eqnarray} 
where $\beta=1$ and $R_2(r,\infty)=
 1- {{\rm sin}^2(\pi r)/ \pi^2 r^2} -
\left(\int_r^\infty {\rm d}x {{\rm sin}\pi x / \pi x}\right)
\left({{\rm d}\over {\rm d}r}{{\rm sin}\pi r / \pi r}\right)$
(GOE limit). 

As can be seen from the above, 
$Y_2 (r,\Lambda) \approx -{\frac {2\Lambda}{4\pi^2\Lambda^2+r^2}} +
{\frac {1}{2\pi^2 r^2}}$ for $ r > {\sqrt\Lambda}$. However, note, 
for $r>2\pi\Lambda$, the behavior of $Y_2$  is different from that 
of a BE with no TRS: $Y_2 \approx {(1-4 \beta \pi^2\Lambda)/2 \pi^2 r^2}$.  
This further suggest following behavior of $I$:
$I=1-(\beta \pi^2 \Lambda)^{-1}$. 
The $I$ for a critical BE is therefore symmetry-dependent; 
 (as $\Lambda=\Lambda_b$ does not depend on $\beta$). 
However the $I$ for its AE analog is independent of the symmetry 
parameter $\beta$; this is because 
$\Lambda =\Lambda_a \propto \beta^{-1}$ in this case (see eq.(19)).   
  
\vspace{.1in}  
{\bf (ii) Nearest Neighbor Spacing Distribution P(S)}

\vspace{.1in}  
The $P(s)$ for this case can be given by using the one for a  
 BE during Poisson $\rightarrow$GOE transition \cite{to}:
\begin{eqnarray}
P(s,\Lambda)=(\pi/8\Lambda)^{1/2} s {\rm e}^{-s^2/16 \Lambda}
I_0(s^2/16\Lambda)
\end{eqnarray}
with $I_0$ as the Bessel function; note, as expected, this is same as $R_2$ 
behavior for small-$r$ (eq.(37)).   

\vspace{.1in}  
{\bf (iii) Level-Compressibility} 
\vspace{.1in}  

The lack of the knowledge of $R_2(r,\Lambda)$ for entire energy-range 
 handicaps us in providing an exact form of the compressibility for the 
time-reversal case.
However its approximate behavior can be obtained by using the relation 
$\chi=1-I$. Thus, for a time-reversal critical BE ($\beta=1$), 
  
\begin{eqnarray}
\chi \approx( \pi^2 \Lambda)^{-1}
\end{eqnarray}
and therefore $\chi \approx ( \pi^2 \Lambda_a)^{-1}$ for its AE analog.

Equations (35, 41) indicate the influence of underlying symmetry on the 
compressibility of the levels: 
$\chi \approx ( \beta\pi^2 \Lambda)^{-1}$. Note $\chi$ for a BE 
is symmetry-dependent due to $\Lambda=\Lambda_b$ being  
$\beta$-independent (see eq.(16)). 
 However as $\Lambda=\Lambda_a \propto \beta^{-1}$ for an AE (see eq.(19)), its $\chi$  
would be symmetry-independent; this is in agreement with numerical 
observations for Anderson systems\cite{hs1,ter,vep}. This further implies that the 
critical BEs corresponding to critical AEs with and without time-reversal 
symmetry would be different. 

	    In past, an attempt to explain the symmetry independence of the 
level-statistics at the Anderson transition was made in the study \cite{vep} by 
suggesting a scaling behavior of the distribution $P(s)$  with the conductance $g$
and the symmetry parameter $\beta$. The $P(s)$ in the study \cite{vep} was obtained 
by interpolation between metallic and Insulator limits.  
We note that, by using the $\Lambda-g$ connection 
(eq.(22)), the $P(s)$ given by eqs.(32,40) can also be 
expressed as a function of $g$.

\section{Numerical Comparison of the Level-Statistics of critical AE 
and critical BE} 

        In this section, we investigate the AE-BE spectral analogy by 
numerically comparing two of their fluctuation measures,
namely, $P(s)$ and the number variance $\Sigma^2(r)=<(r-<r>)^2>$. 
The former is a measure of the short-range correlations in the 
spectrum and the latter, describing the variance in the number of 
levels in an interval of $r$ mean level spacings, contains the 
information about the long-range correlations \cite{me}.    
The $\Sigma_2$ is also an indicator of the compressibility of the 
spectrum; ${\rm lim}\;  r\rightarrow \infty \; 
\Sigma_2(r) \approx \chi r$. To study the AE-BE analogy in presence of time-reversal 
symmetry as well as its absence, we consider two cases of the 3-dimensional AE 
(simple cubic lattice of size $L=13$ and with Gaussian site disorder) in critical 
regime:

(i) {\bf AE$_t$ :} 
                                                                                                
The AE with isotropic random hopping, hard wall boundary conditions
    and time-reversal symmetry; here $W=4.05$, $W_1=1$, $W_2=0$,
    $t_1=0$, $t_2=0$.
    The criticality of AE for same disorder parameter values but with
    periodic boundary conditions is numerically studied in \cite{ro}.
    However the system remains in the critical regime under hard wall
    boundary conditions too.

(ii) {\bf AE$_{nt}$ :} 
    The system {\bf G} with isotropic non-random hopping, periodic boundary conditions 
    and no time-reversal symmetry; here $W=21.3$, $W_1=0$, $W_2=0$, 
    $t_1=1$, $t_2=1$. The time-reversal symmetry is broken by 
applying an Aharnov Bohm flux $\phi$  which gives 
rise to a nearest neighbor hopping $H_{kl}={\rm exp}(i \phi)$ 
for all $k,l$ values related to the nearest-neighbor pairs \cite{ter}.   
The flux $\phi$ is chosen to be non-random in nature, that is, 
$<{\rm cos}^2(\phi)>=W_1=0$, $<{\rm sin}^2(\phi)>=W_2=0$ and  
$<{\rm cos}(\phi)>=t_1=1$, $<{\rm sin}(\phi)>=t_2=1$.

We study each AE case for two system sizes  $L=10$ and $L=13$ 
by numerically diagonalizing the matrices of the ensembles by standard techniques. 
Each ensemble contains few thousand matrices and the statistical average is performed 
approximately over $3\times 10^5$ levels, obtained by taking 200 levels in a small 
energy range around center 
$E=0$ of the spectrum of each matrix. Each BE (chosen with $\gamma=2$) is also analyzed 
for two dimensions $N=1000$ and $3000$. Note, due to a higher rate of change of 
the mean level density, the $\Lambda$ (eq.(16)) for BEs changes more rapidly with energy  
as compared to AE cases (eq.(19)). To avoid mixing of levels with 
different transition rates, therefore, fewer ($\approx 100$) levels are taken from the 
spectrum of a matrix in the BE case; 
the total number of levels for BE analysis is kept nearly same (as in the AE cases) 
by taking a bigger ensemble. Each spectrum is unfolded for $P(s)$ and 
$\Sigma_2(r)$-analysis. The unfolding is carried out by numerically 
calculating the unfolded levels $r_j = \int_{-\infty}^{E_j} {\rm d}x R_1 (x)$ 
for ($j=1,2,..,N$) with symbol $E_j$ used for levels before unfolding.

The parameter $Y-Y_0$  for both AE cases  is given by eq.(5). As  $N|Y-Y_0| \simeq o(1)$,  
the mean level-density $R_1$ for both AE cases is given by eq.(14); 
$R_1=N F$ with $F$ as an $N$-independent function of energy. This is confirmed 
by our numerical analysis of $R_1$ for different $N$-values for each AE case,  
with  figure 2(a) showing the comparison only for two 
$N$-values. (Note, for $R_1$ study, the spectrum is not unfolded and almost all eigenvalues 
of each matrix are used for the analysis). As mentioned in section III,  the critical BE analogs 
for the fluctuations measures need not have a same $R_1$. 
The $R_1$-behavior for the critical BE analogs for the fluctuations of the 
AE cases is shown in figure 2(b); the numerical fitting confirms that 
$R_1= (N/\sqrt{\pi}) {\rm e}^{-E^2}$ for each critical BE case which is quite 
different from their AE analogs.     

The figures 3 and 4 show the $P(S)$ and $\Sigma_2(r)/r$ behavior for the two AE cases. 
The almost same behavior for two system sizes in each AE case, for both the measures,  
confirms their critical nature. We find, from figure 4, that  the large-$r$ behavior of 
$\Sigma^2(r)/r$ for both AEs  seems to converge to 
$\chi \approx 0.25$ which confirms our analytical prediction  of $\chi$ for AEs 
(with $\zeta \sim o(L^{D_2/d})$,based on equality of eqs.(25,27)); it is also in 
agreement with other numerical studies  \cite{bran,ckl,ro1,ter}.

	The determination of the critical BE analog of the fluctuation 
measures of an AE requires a prior information about $\Lambda_a$ (given by eq.(19)). 
Although we know the function $F$ for each AE case (see figure 2(a)) as well as 
$\alpha-\alpha_0$ (from eq.(6), $\alpha-\alpha_0= 1.36, 5.43$  for $AE_t$ and $AE_{nt}$, 
respectively), however the determination of $\zeta$ requires an  statistical analysis 
of wavefunctions. Fortunately equations (35,41) suggest that the 
parameter $c$ of the critical BE analog of a critical AE can also be obtained 
(approximately) from its $\chi$ behavior: 
$c \approx (\beta\pi \chi/ 4\gamma)$. Using $\chi \approx 0.25$, the theoretically expected $c$ 
parameters for $AE_t$ and $AE_{nt}$ are $0.1$ and $0.2$, respectively. 
The figures 3 and 4 confirm the existence of the critical BE analogs, of the AE cases, 
at the above $c$ values. Note the above relation between the 
parameter $c$ of a critical BE and $\chi$ of a critical AE is obtained by  combining the 
theoretical results for (i) critical 
point $D_2$ behavior predicted by our $\Lambda$ formulation, (ii) $D_2$ given by eq.(27), 
 (iii) $\chi$ for a BE (eqs.(35,41)), (iv) AE-BE analogy. The good AE-BE agreement observed 
in figure 3, 4 therefore indicates the validity of all the above formulations, used to 
derive $c(BE)-\chi (AE)$ relation.

	The exponential decay of the tail of $P(s)$ for $d>2$ AE system at the critical 
point has been confirmed by various numerical studies (for example, see 
\cite{ss,bran,ckl,ro1,ter,sp}). The validity of AE-BE analogy 
requires a similar decay of $P(s)$ for the critical BEs too. 
The figure 1(b) compares $P(s)$ behavior for a few critical BEs with  
the function ${\rm e}^{-\kappa s}$. The fitted $\kappa$ values are 
close to  $\kappa\approx (\beta \pi /8 \gamma c)$ for 
intermediate $c$ ranges; the $\kappa - c$ relation 
is obtained by using $\kappa=(2\chi)^{-1}$ and $\chi=(4\gamma c/\beta \pi)$ at $E=0$ 
(see below eqs.(35)). The insets in figure 3  compare 
the tails of the $P(s)$ for the AEs and their BE analogs with function 
${\rm e}^{-\kappa s}$; we find $\kappa\approx 1.5 -1.7$. The result is close 
to our analytical prediction $\kappa \approx 2$ for the critical AE case (see below 
eq.(27)). However the lack of exact agreement seems to suggest the 
approximate nature of the $\chi$-$\kappa$ 
relation, namely, $\kappa=(1/2\chi)$; (note as our analytical prediction 
$\chi=4\gamma c/\beta \pi$ is found to be in excellent agreement with numerics, 
this leaves $\chi-\kappa$ relation as the possible source of error).

.   
	The study \cite{km} claims that the critical level-statistics in the 
Rosenzweig-Porter ensemble (similar to BE, as mentioned in section II) does not 
have a fractional compressibility and, therefore, is different in nature from 
that for critical AE. However our analytical results, supported by 
the numerical evidence, disprove their claim. 
Our numerical study  confirms the 
 existence of a fractional $\chi$, increasing with $c$, for various critical BEs. 
Two such cases are shown in figure 4, with their $\chi$-results in close 
agreement with our analytical prediction, namely, eqs.(35, 41).   

\section{Connection with PRBM model}

	In past, a random matrix ensemble, namely, power law random banded 
matrix (PRBM) ensemble was suggested as a possible model for the critical 
level statistics of Anderson Hamiltonian \cite{mir}. A PRBM ensemble is defined 
as the ensemble of random Hermitian matrices with matrix elements 
$H_{ij}$ as independently distributed Gaussian variables with zero 
mean i.e $<H_{ij}>=0$ and the variance 
$<H_{ij;s}^2>= G_{ij}^{-1} \left[ 1+ (|i-j|/b)^2 \right]^{-1} $, 
$G_{ij}=\beta (2-\delta_{ij})$ and 
$G_{ij}=1/2$. It is critical at arbitrary values of the parameter $b$ 
and is believed to show all the key features of the Anderson critical 
point, including multifractality of eigenfunctions and the 
fractional spectral compressibility.  

The success of  PRBM ensemble, a one dimensional system, as a model for 
Anderson systems 
in arbitrary dimension is a little surprising. However, it 
can be explained on the basis of our formulation.
The PRBM-AE connection is a special case of our study connecting  
any generalized Gaussian ensemble with BE. The PRBM ensemble  
being  Gaussian in nature, its complexity parameter can be defined by 
using eq.(4) which can then be used to obtain its BE analog. 
The eq.(4) gives (with $\gamma=\beta$) 
\begin{eqnarray}
Y-Y_0 &=& {1\over N(N+2-\beta)} 
\sum_{r=1}^N (N-r) {\rm ln}|1+(b/r)^2| \nonumber \\
\end{eqnarray}
which gives $Y-Y_0 \propto  {f(b) / N}$ with $f(b)\approx 2 b^{0.85}{\rm ln}(5 b)$ 
for $b >> 1$ and $f(b) \approx 2 b^{1.75}$ for $b << 1$. 
As $Y-Y_0 \approx o(1/N)$, the $R_1$ can then be given by eq.(14).  
Following eq.(15), $\Lambda$ for a PRBM ensemble is
\begin{eqnarray}
\Lambda_{prbm}(b,E) = f(b) F^2(E) \zeta^2 N^{-1}. 
\end{eqnarray}

	The well-known size independence of level-statistics for PRBM case for all 
$b$ values requires $\Lambda_{prbm}$ to be $N$ independent which gives 
$\zeta \propto N^{1/2}$ for all $b$ ranges by our formulation. 
However note that, for PRBM model, $<I_2> \propto N^{-D_2}$ with $D_2$ as a function 
of $b$\cite{em,ver,cu1}. The use of $\zeta \propto <I_2>^{-1}$ in eq.(43), therefore,  
gives a size-dependent $\Lambda_{prbm}$. Keeping in view the well-known criticality 
of PRBM system for all $b$ ranges,  it seems that the relation $\zeta \propto <I_2>^{-1}$  
is not valid for the PRBM case.

	Using the prediction $\zeta =\zeta_0 N^{1/2}$ in eq.(43) and  the 
relation $\Lambda_{prbm}=\Lambda_b$, the level-statistics of a PRBM can be mapped to a 
critical BE ensemble with 
\begin{eqnarray}
c=(4\pi \beta \Lambda_{prbm})^{-1} {\rm e}^{-E^2}
=(4 \pi \beta f(b) \zeta_0^2 F^2(E) {\rm e}^{E^2})^{-1}
\end{eqnarray}

The spectral statistics of PRBM therefore shows a crossover from 
from Poisson  (as $c\rightarrow \infty$ for $\Lambda_{prbm}\rightarrow 0$ 
i.e $b \rightarrow 0$) to Wigner-Dyson behavior ($c\rightarrow 0$ for 
$\Lambda_{prbm}\rightarrow \infty$ or $b \rightarrow \infty$).    

The spectral compressibility $\chi$ for a PRBM ensemble at $E=0$  can now be obtained 
by substituting $\Lambda=\Lambda_{prbm}= f(b) \zeta_0^2$ in the 
eqs.(34,35,41) which give
\begin{eqnarray}
\chi &=& 1 - 4 \pi^2  \zeta_0^2 f(b) \qquad {\rm for}\; b << 1 \nonumber \\ 
&=&  (\beta \pi^2 \zeta_0^2 f(b) )^{-1}  \qquad {\rm for} \;  b >> 1. 
\end{eqnarray}
The above results are at least in the same form as obtained 
in \cite{em,ver,cu1}; the lack of explicit knowledge of $\zeta_0$ prevents 
us from making any further comparison. As $\chi$, in the eq.(45), changes  
from $0$ to $1$  with decreasing $b$, it violates 
the relation (27) in the range $b<<1$. A same violation was observed in previous
PRBM studies \cite{em} too. Thus our results obtained by using PRBM-BE analogy 
seem to be in accordance with earlier studies on PRBM model.

        In brief, the PRBM ensemble, with $b$ as a parameter, can be mapped to 
critical BE with parameter $c$ (see eq.(44)). As a consequence, the studies 
suggesting the analogy of spectral statistics for the PRBM and AE ensembles are 
well in agreement with our study claiming the  AE-BE analogy. By using the    
connection of PRBM with MNS model \cite{km,mns}, the PRBM-BE-AE analogy can further 
be extended to MNS-BE-AE analogy.

\section{conclusion}

	In the end, we re-emphasize our main result:

Under independent electron approximation, the level-statistics for 
the disordered systems undergoing localization $\rightarrow$ 
delocalization transition of wavefunctions can 
be described by the Brownian ensembles (with uncorrelated elements) 
undergoing a similar transition. 

	The analogy helps us in making following deductions:



	(i) The transition in the statistics is governed by a 
single scaling parameter $\Lambda={|Y-Y_0|\over \Delta^2} ({\zeta\over L})^{2d}=
f({\zeta\over L})$. The second equality follows from the 
dependence of wavefunction statistics e.g inverse participation ratio $I_2$ and 
therefore $\zeta$ on the complexity parameter $|Y-Y_0|$.

(ii) The level-statistics is governed by 
the competition between complexity parameter and local mean level 
spacing. The critical point of level statistics occurs when the 
      complexity parameter $Y-Y_0$ and $\Delta_l$ have same size 
dependence. In particular, if $|Y-Y_0| \sim o(N^\alpha)$ and 
$\Delta_l \sim o(N^{\beta})$ for a disordered system, its critical point 
will occur when $\alpha-2\beta =0$.   However if the local mean level 
spacing in the system changes at a faster rate with size as compared to 
$\sqrt{Y-Y_0}$ (i.e $\beta > \alpha/2$), the system will never reach 
its critical state and will always remain in the localized regime.

(iii) The critical BE analog of a critical AE is unique. 
      Further, it is different for critical AE with and without 
      time-reversal symmetry. Similar to AEs, the level-statistics of BEs 
shows a scaling behavior 
as well as a critical point with fractional level-compressibility. 
However, unlike AEs, the $\chi$ turns out to be symmetry dependent for BEs, 
their parameter $\Lambda$ being symmetry independent.

 (iv)  The AE-BE analogy confirms the symmetry-independence of the 
compressibility of levels and the multifractality of the wavefunctions 
at the critical point of Anderson transition. The analogy also indicates  
the non-validity of relation $D_2=d(1-2\chi)$ in the strong multifractality 
limit, and, the approximate nature of the relation $\kappa=(2\chi)^{-1}$. 

 (v) The AE-BE analogy  helps us in formulating, for the first time, the exact 
2-point level density correlation at the critical point of a disordered 
system. The formulation is applicable for a wide range of system parametric 
conditions.

 	It should be noted that both MIT as well as WDT occur due to 
delocalization of the wavefunctions. In fact, our analytical study 
suggests that the level-statistics of almost all complex systems undergoing a 
localization $\rightarrow$ delocalization 
transition follows the same route although with different transition 
rates; the state of level-statistics of two systems with different 
complexity may correspond to two different points on this route.
In principle, our analytical work is applicable to the Gaussian models of 
complex systems only, however the intuition based on earlier studies suggest the 
validity of the results for the systems with other origins of randomness too 
\cite{me}. For example, the investigation of a number of dynamical systems seems to 
support this intuition. It has been shown that the spectral statistics of pseudo-integrable 
billiards is remarkably similar to the critical statistics of AE \cite{bo}. The presence 
of a statistics intermediate between the Poisson and GOE has already been shown for 
the Kicked rotor in the non-integrable regime of the kicking parameter 
\cite{ds}. A correspondence  of the integrable systems to the insulators 
and of the chaotic systems to the metals is already known to exist.
The Integrability $\rightarrow$ chaos transition in the dynamical 
systems therefore seems to follow a route in the level-statistics similar to 
that of MIT; note such a transition in classical systems corresponds to a 
delocalization of the wavefunctions in their quantum analog.     
Thus the analogy of the statistical level fluctuations between AE and BE may possibly 
be extended to dynamical systems and BE too; if the latter is found to be correct, 
the analogy would be useful to obtain the correlations for the non-integrable regime.  
	
	The evidence of such an analogy would suggest the existence of 
several features, unknown so far, for the level statistics of dynamical 
systems. For example, the analogy can be used to intuitively claim and search 
for the existence of a critical point, the dimensional dependence of level-statistics 
and the multifractality of eigenfunctions during the transition from Integrable to 
chaotic dynamics. 
It should be noted that a generic one-dimensional dynamical system always 
shows a Poisson level-statistics (in analogy with one-dimensional AE). However 
the dynamics in a 3-dimensional system shows a feature namely "Arnold Diffusion", 
absent in lower dimensions. The intuition based on the above analogy 
suggests the possible existence of a critical level statistics at the parametric 
values at which Arnold Diffusion takes place. A further exploration of such an
analogy is therefore highly desirable.

\section*{Figure Captions}
\noindent Fig. 1. The study of critical BEs:

(a) The scaling behavior of the integrated nearest-neighbor 
spacing distribution $\alpha$ during WDT. Note $\alpha$ values for  
BEs with different parameters $\mu$ and sizes $N$ converge on the 
same two curves, thus indicating $\alpha$ dependence on a 
specific combination of $\mu$ and $N$, namely, $z=\mu/N^2$. 
Further, at $z=c$, $\alpha$  remains unchanged for different $N$ values, 
thus indicating a critical point of BEs.

\vspace{.05in}

 
(b)  The comparison of tail of $P(S)$ distribution for two of the critical BEs with 
function ${\rm e}^{-\kappa s}$ for $\beta=2$ case. For intermediate $c$-ranges, 
the $P(s)$ is well-fitted by the function ${\rm e}^{-\kappa s}$. The fitting 
however, seems to be poor for smaller $c$ values which is as expected, due to 
statistics approaching GUE limit (which corresponds to $P(s) \sim {\rm e}^{-\pi s^2/4}$). 
The fitted $\kappa$-values are as follows:
(i)$\kappa=0.8$ for $c=0.3$,
(ii) $\kappa=1.7$ for $c=0.03$.
 The above $\kappa$-values seem to deviate significantly from the relation 
 $\kappa=(\pi/4\gamma c)$ (obtained by using $\kappa=(2\chi)^{-1}$ with 
$\chi$ given by eq.(35) and $\gamma=2$). We have seen a similar deviation for 
the BEs  with $\beta=1$ too. This suggests the non-validity 
of relation $\kappa=(2\chi)^{-1}$ in general although it seems to work for 
some $c$ values (see for example figure 3).

\vspace{.1in}

\noindent Fig. 2.  The behavior of level-density $F(E)=N^{-1}.R_1$: 
\vspace{.05in}

(a) for the two cases for two system sizes $L=10$ and $L=13$. 
The numerical fitted function has the form 
$F=f_1 {\rm e}^{-(E^2/f_2)}$ with $f_1=0.16, f_2=5$ for AE$_{t}$ and 
$f_1=0.016, f_2=400$ for AE$_{nt}$.

\vspace{.05in}

(b) for the critical BE analogs of the higher-order correlations of the two cases 
considered in 2.(a). Here $F$ for all the critical BE cases is 
well-fitted by the function $F(E)=\pi^{-1/2} {\rm e}^{-E^2}$. 
Note the lack of analogy between the mean-level densities 
for the cases given in 2(a) and 2(b) while their higher order correlations 
(shown in Figures 3,4) are approximately same. 

\vspace{.1in}
 
\noindent Fig. 3.  The comparison of distribution $P(S)$ of the 
nearest-neighbor spacings $S$ for the AE (d=3) and BE cases on a 
log-log scale. To confirm the critical state of the AE, the distribution is 
shown for two 
system-sizes $L$ for each AE case. The insets show the same functions 
on a lin-log scale and also compares the behavior with ${\rm e}^{-\kappa s}$:
 
(a)  AE$_{t}$ (with hard wall boundary conditions, random hopping and time-reversal 
symmetry) and its  critical BE analog ($c=0.1$). The dashed line in the inset is 
fitted function $f=4 {\rm e}^{-1.7 S}$ which gives $\kappa \approx 1.7$.
(b) AE$_{nt}$ 
 (with periodic boundary conditions, non-random hopping and no time-reversal symmetry)  
along with its critical BE analog $c=0.2$.The dashed line in the insert is fitted 
function $f=2 {\rm e}^{-1.5 S}$ which gives $\kappa \approx 1.5$.

\vspace{.1in}

\noindent Fig. 4.  The comparison of the $\Sigma^2(r)/r$-behavior for 
the AE and BE cases: 
 (a) AE$_{t}$  and the corresponding critical BE analog ($c=0.1$).
 (b) $AE_{nt}$ and corresponding critical BE ($c=0.2$)

Note here the critical BE analog for each AE case is 
same as for the $P(S)$-study.  As can be seen, $\Sigma^2/r$ for large $r$ 
seems to approach the limit suggested by the relation $\chi=
{\matrix{{\rm lim} \cr  r\rightarrow \infty }}\Sigma^2/r = (4\gamma c/\beta \pi)$, 
that is, $\chi=0.25$. Note this is the expected $\chi$ for AEs on the basis of 
eqs.(25,27) too. Besides showing AE-BE analogy, the figure also confirms 
that (i) symmetry independence of $\chi$ for AEs (ii)   
fractional $\chi$-result for a critical BE.




\begin{figure}
\begin{center}
\resizebox{150mm}{!}{\includegraphics{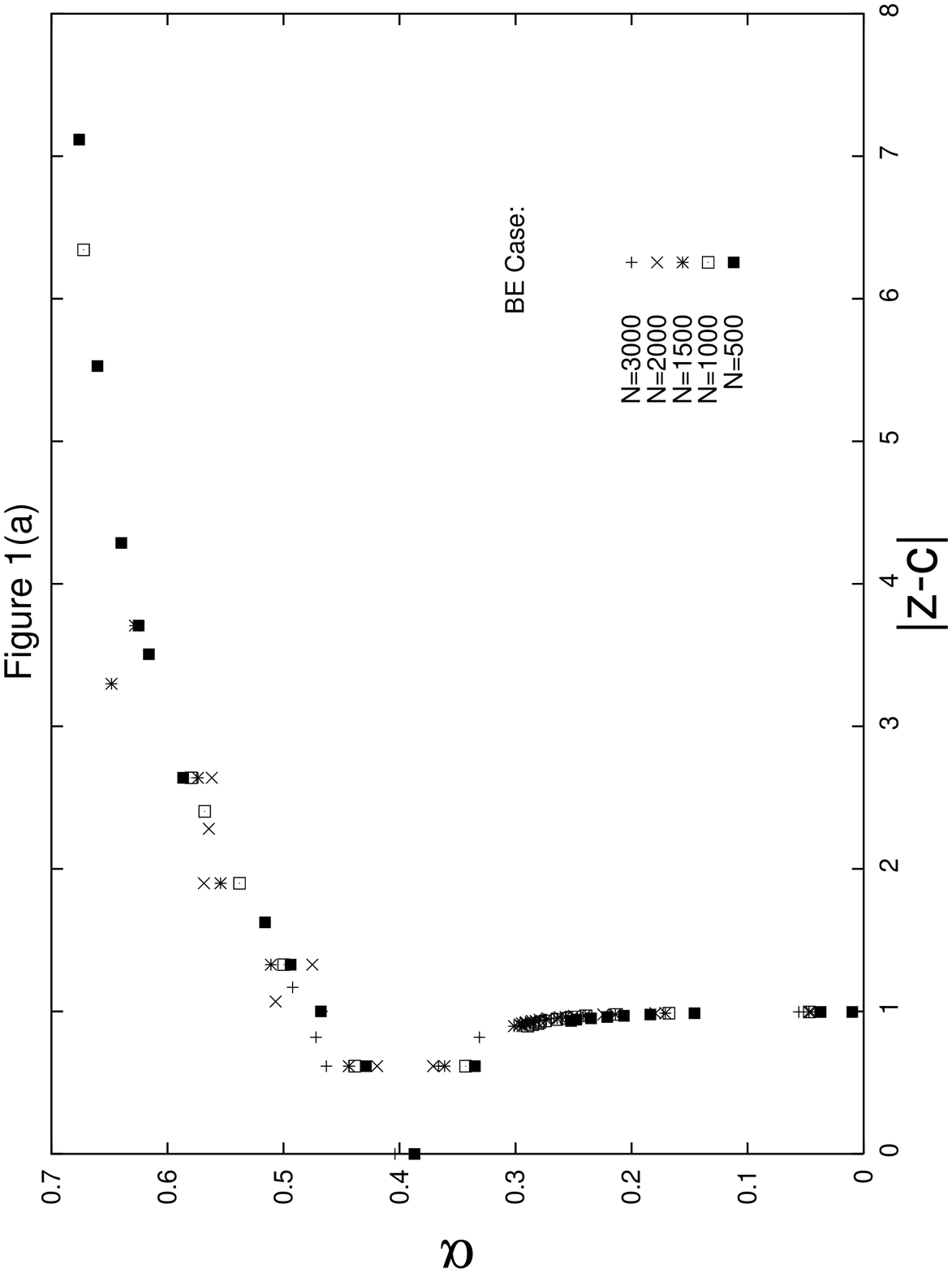}}
\end{center}
\end{figure}

\begin{figure}
\begin{center}
\resizebox{150mm}{!}{\includegraphics{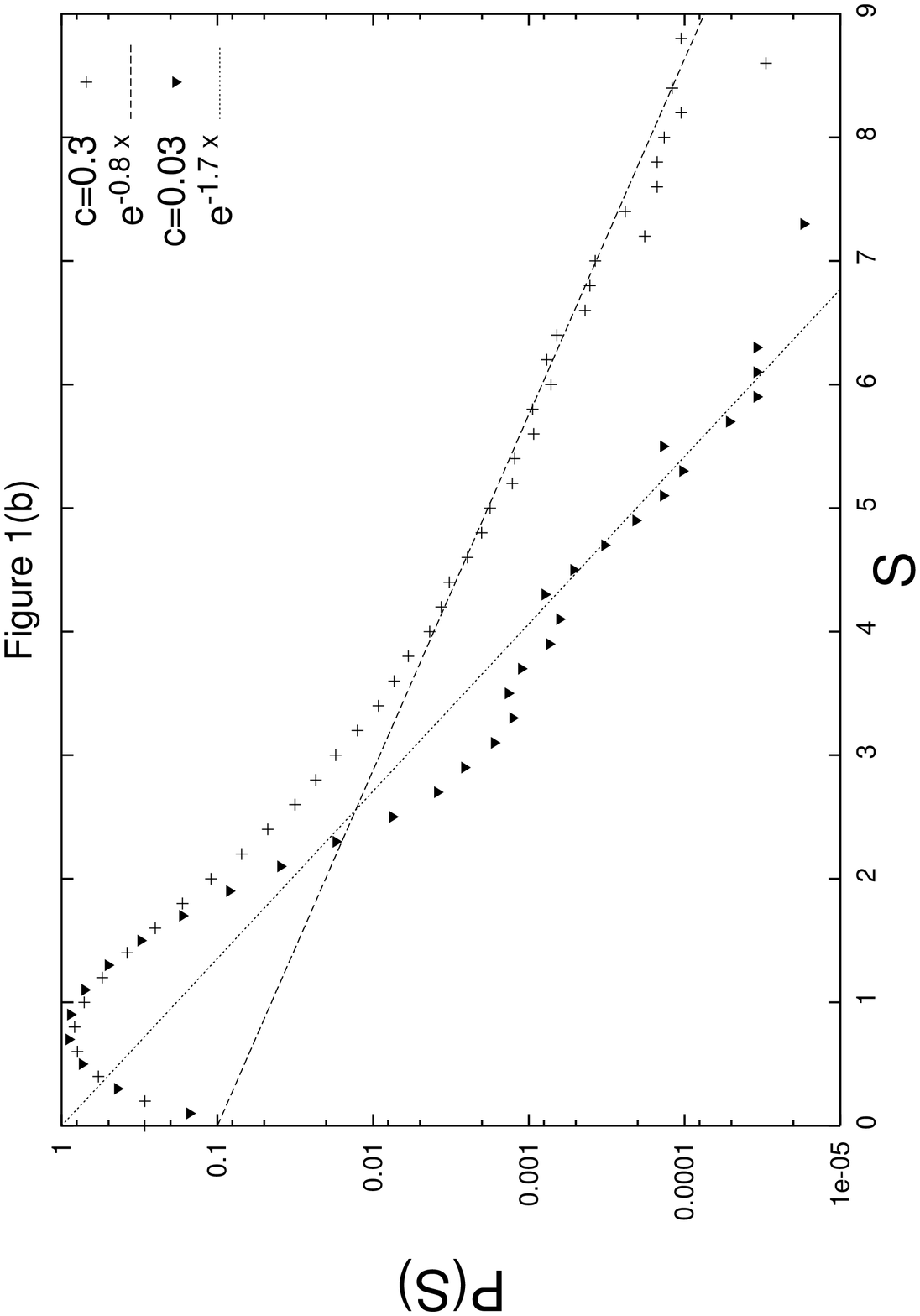}}
\end{center}
\end{figure}

\begin{figure}
\begin{center}
\resizebox{150mm}{!}{\includegraphics{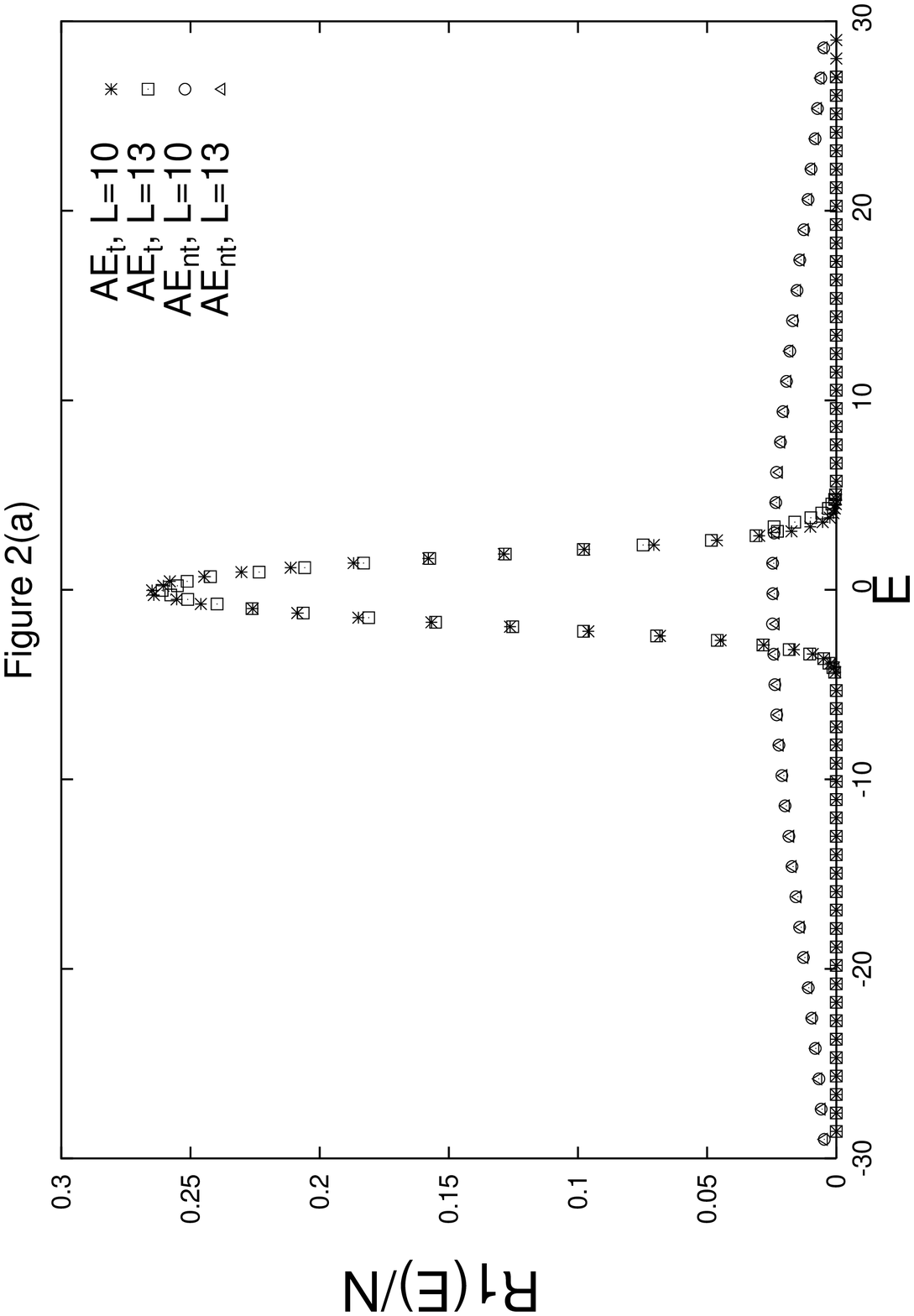}}
\end{center}
\end{figure}

\begin{figure}
\begin{center}
\resizebox{150mm}{!}{\includegraphics{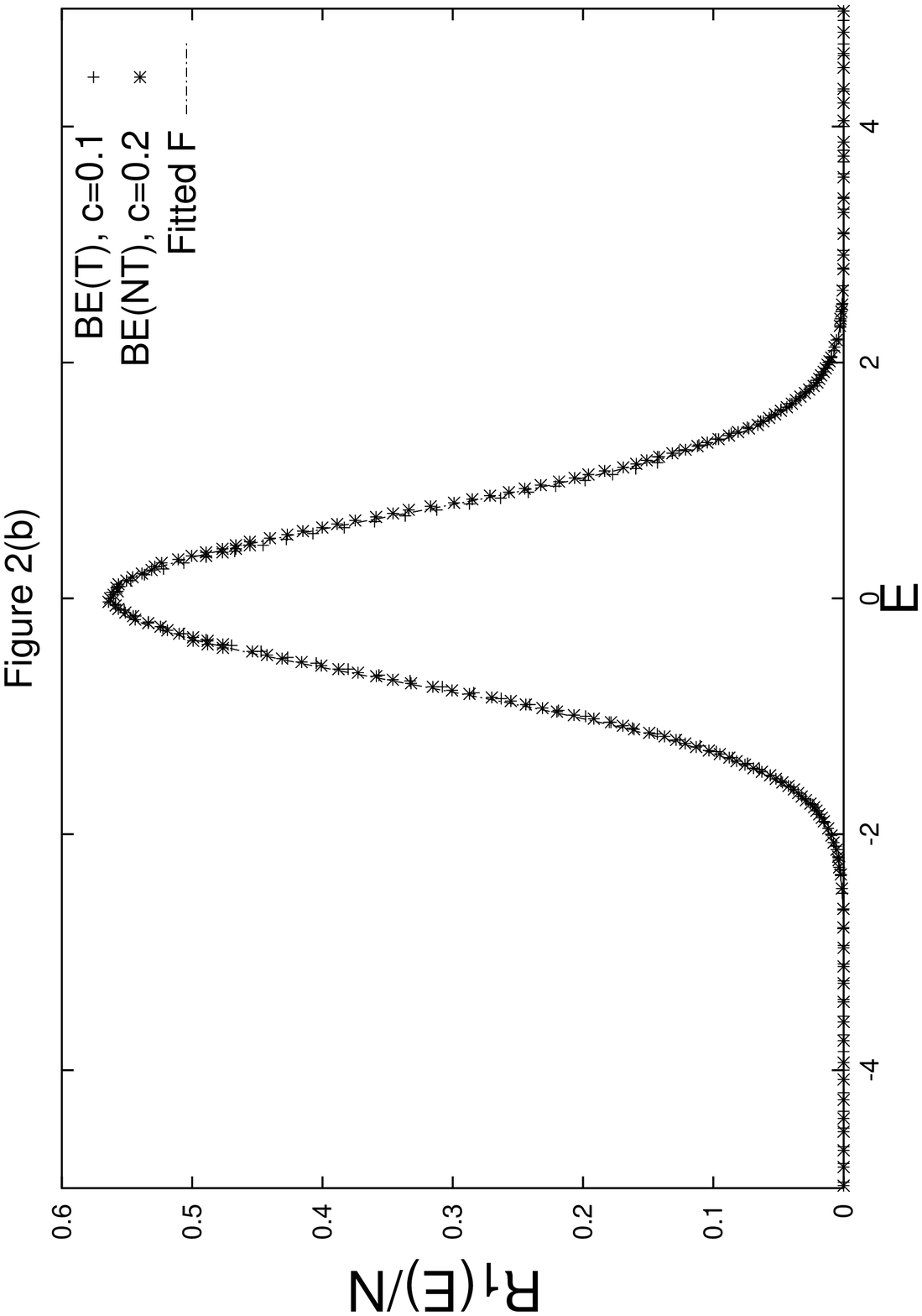}}
\end{center}
\end{figure}

\begin{figure}
\begin{center}
\resizebox{150mm}{!}{\includegraphics{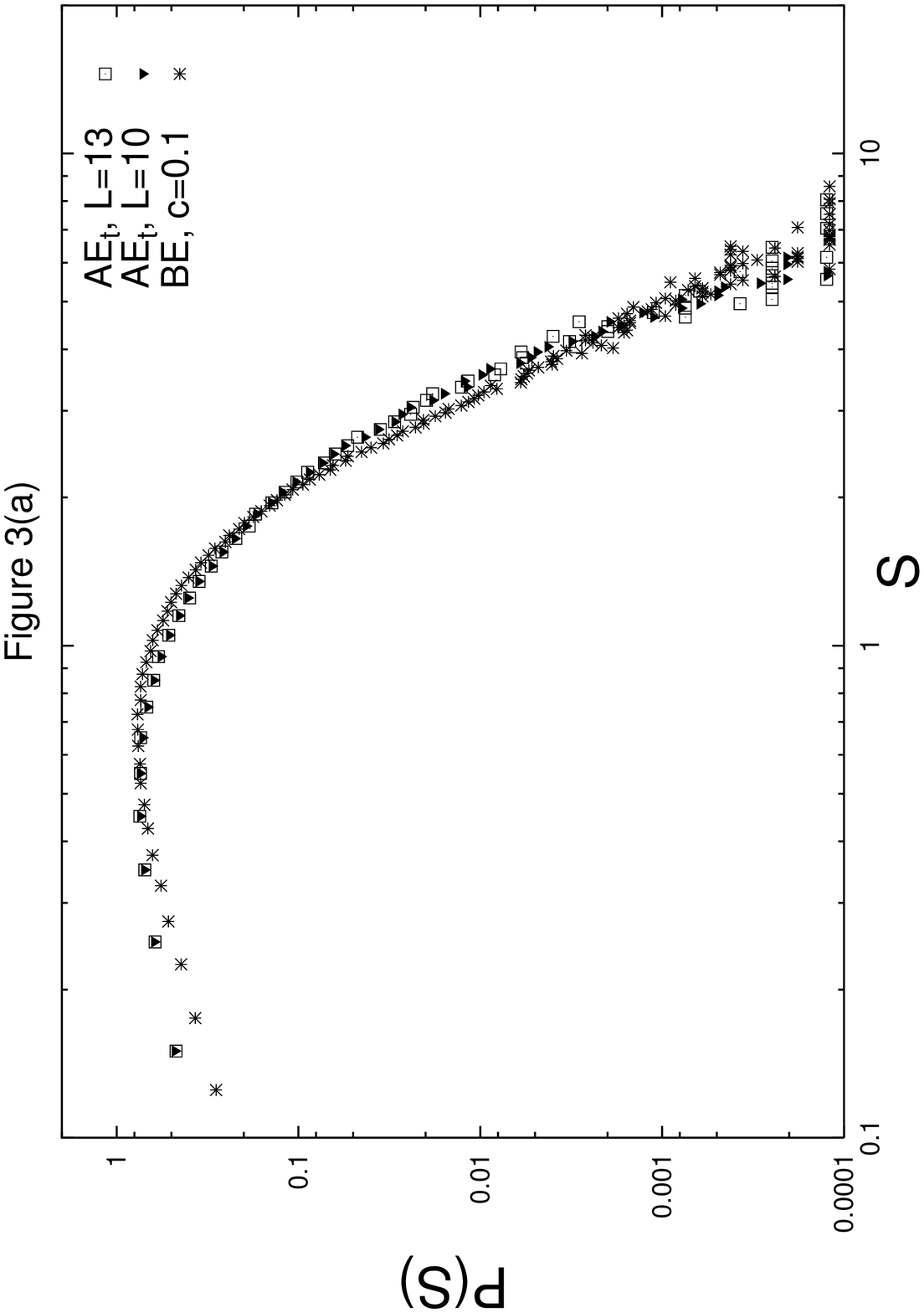}}
\end{center}
\end{figure}

\begin{figure}
\begin{center}
\resizebox{150mm}{!}{\includegraphics{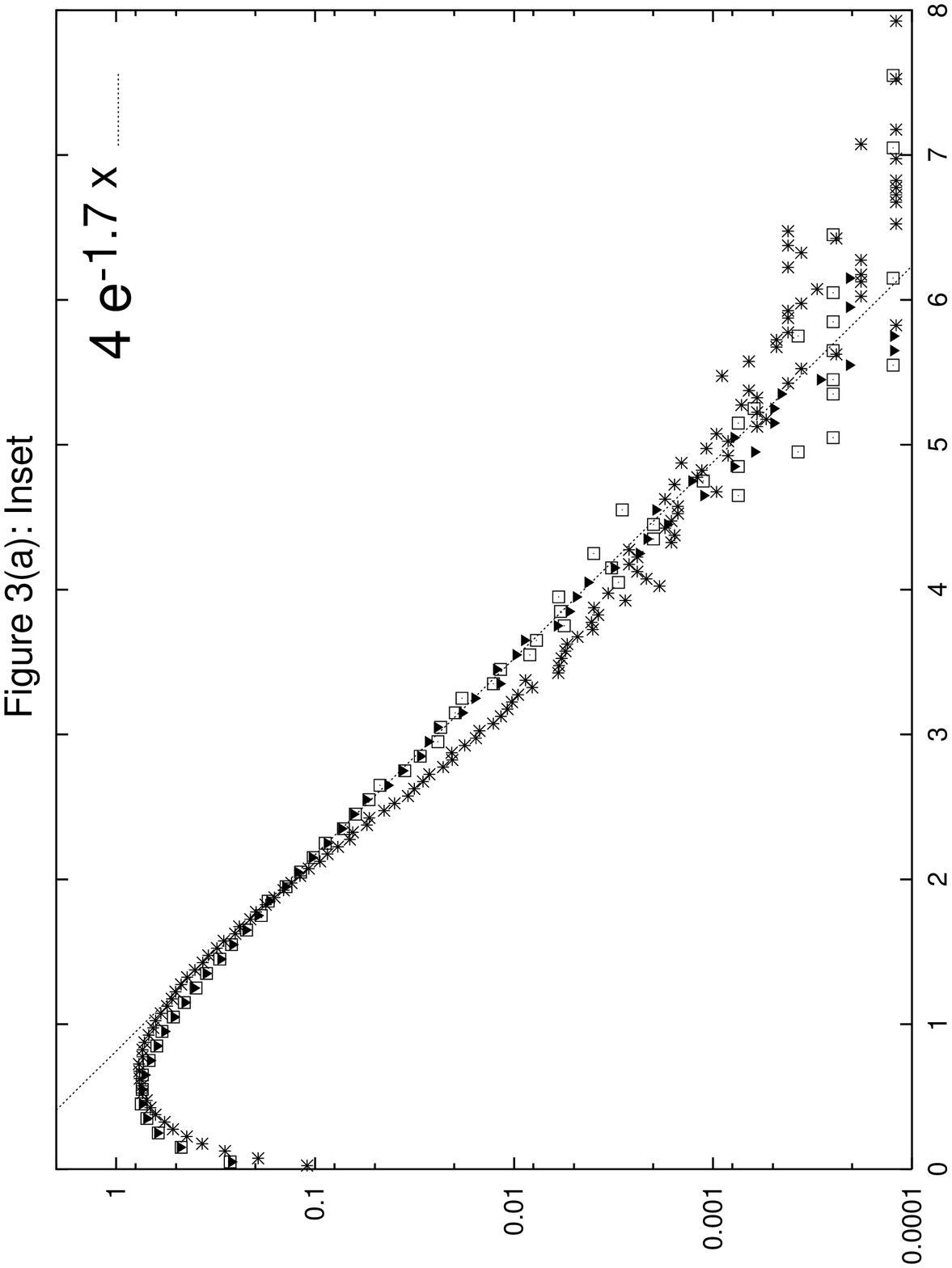}}
\end{center}
\end{figure}

\begin{figure}
\begin{center}
\resizebox{150mm}{!}{\includegraphics{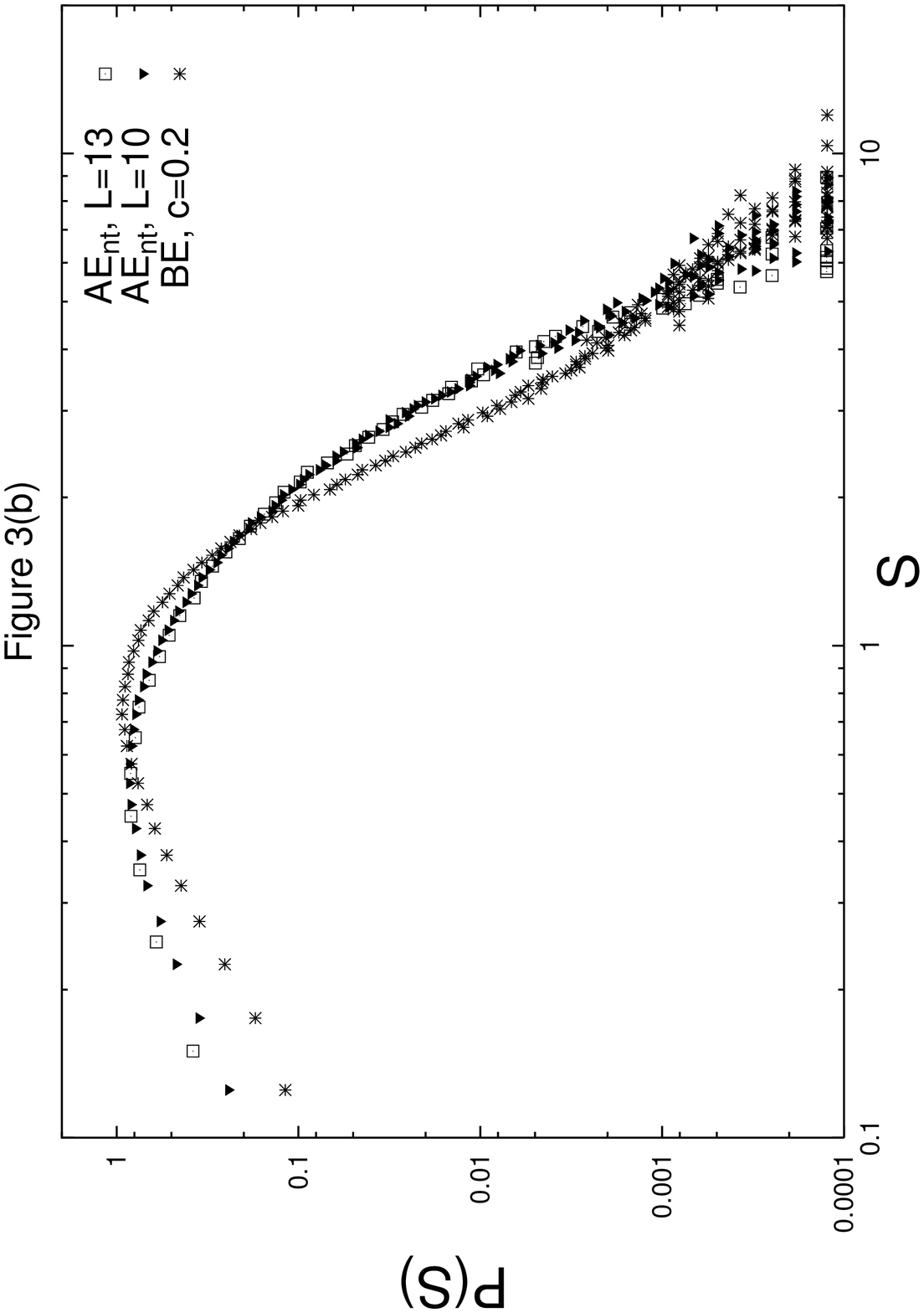}}
\end{center}
\end{figure}

\begin{figure}
\begin{center}
\resizebox{150mm}{!}{\includegraphics{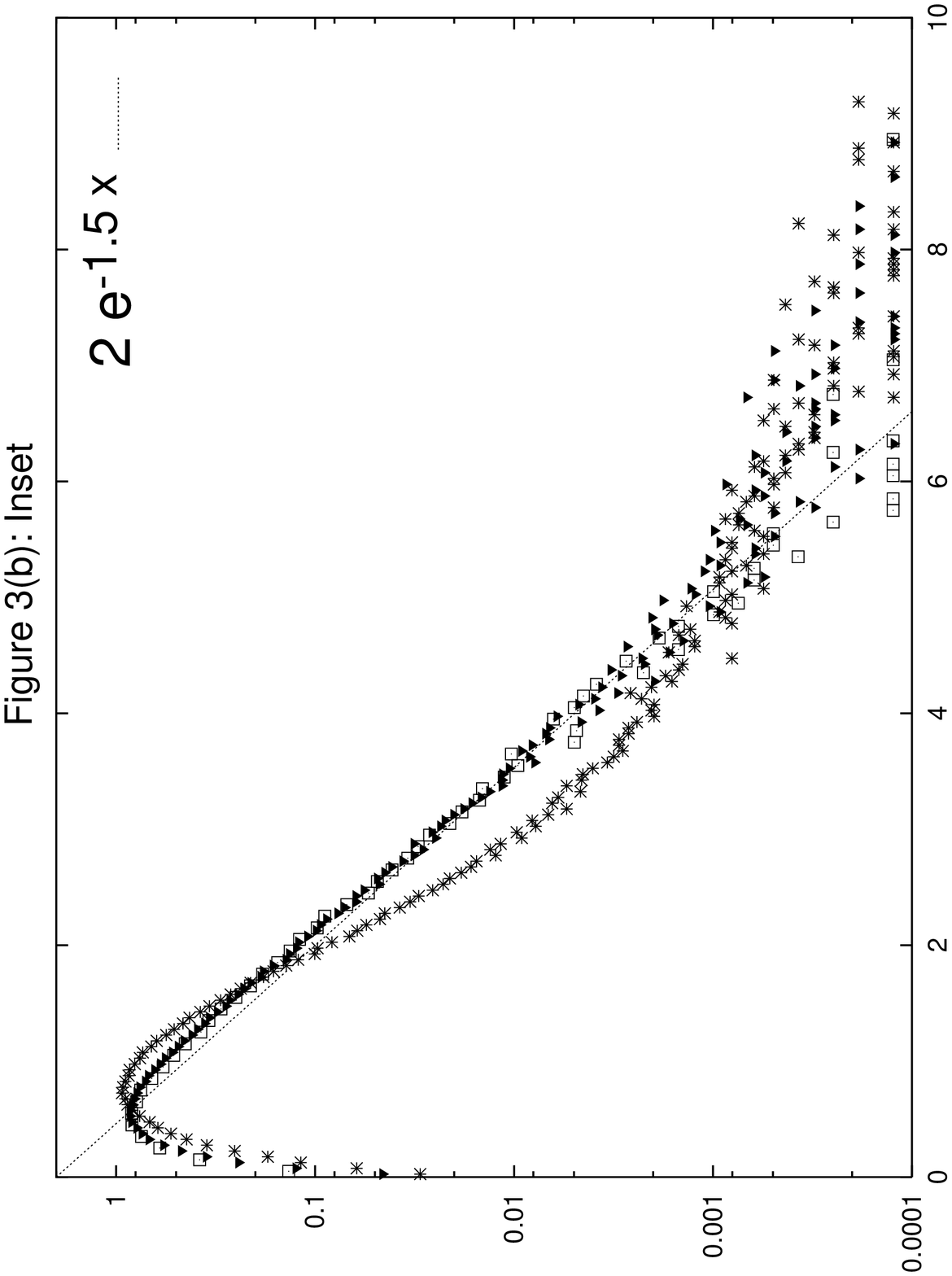}}
\end{center}
\end{figure}

\begin{figure}
\begin{center}
\resizebox{150mm}{!}{\includegraphics{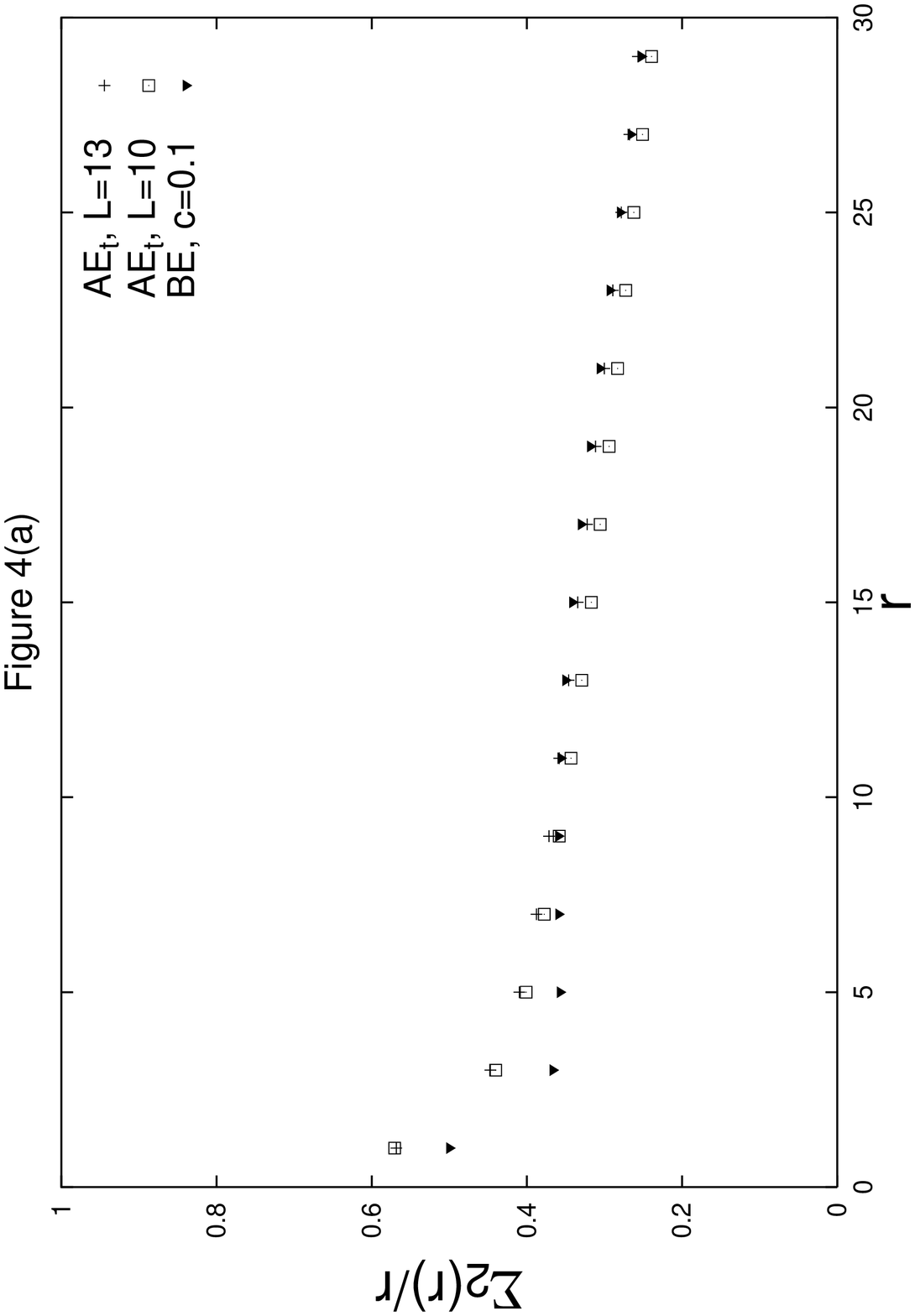}}
\end{center}
\end{figure}

\begin{figure}
\begin{center}
\resizebox{150mm}{!}{\includegraphics{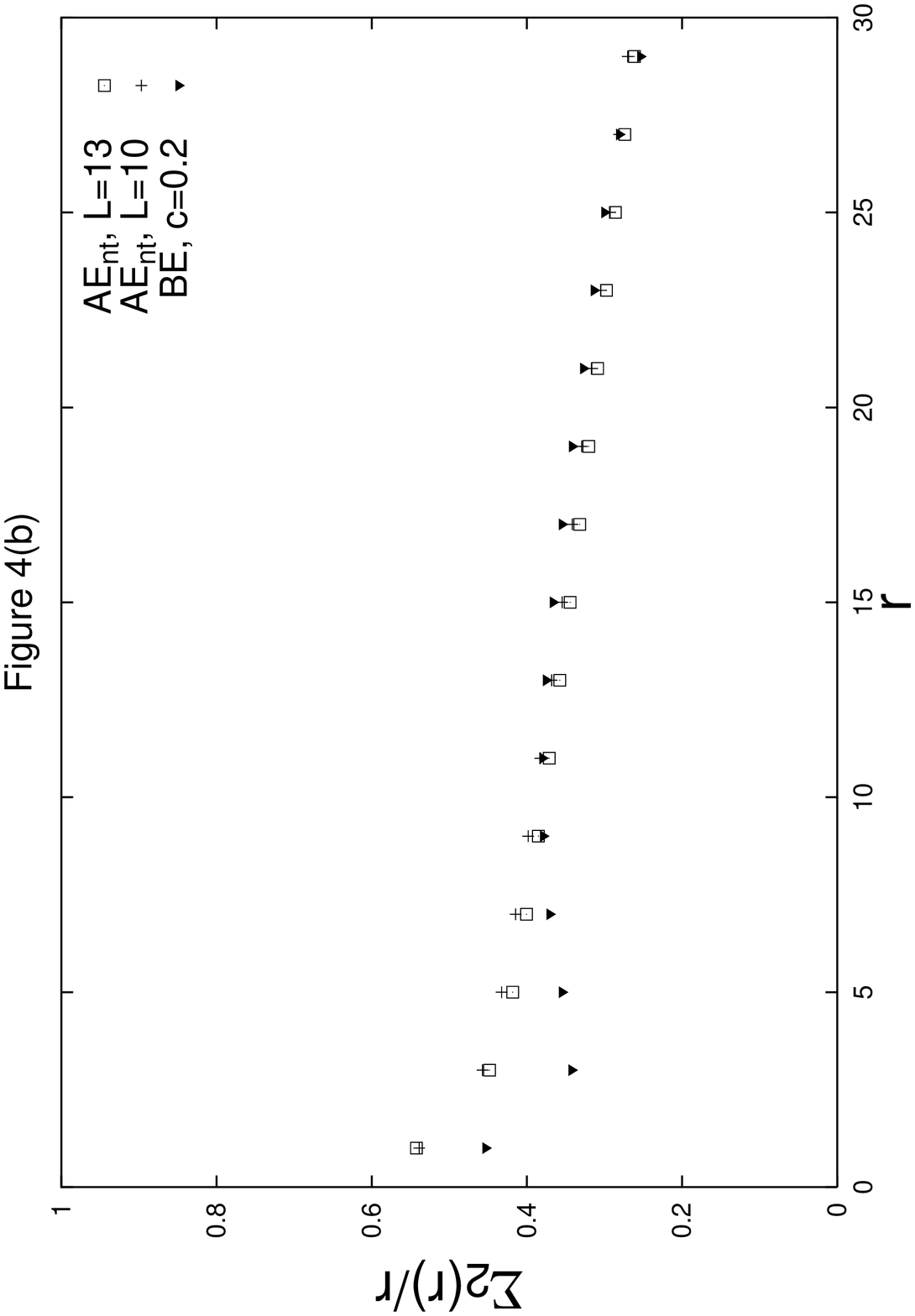}}
\end{center}
\end{figure}


\begin{references}
\bibitem{pp}
The higher order correlations basically being the measures of the 
fluctuations of the density around its average value, their comparison in 
two different spectrum makes sense only if the fluctuations are 
measured with respect to same background. This requires an unfolding 
of each spectrum, that is, rescaling by its mean level density before 
comparison with another spectrum \cite{ap}. As the parameter governing 
the evolution in the rescaled spectrum is $\Lambda$, the higher order 
correlations of an AE are given by a BE with a same $\Lambda$ value.
\bibitem{pp1}
For a $d$ dimensional disordered system, 
the number of states in a volume of linear dimension $\zeta$ in 
$d$-dimensions is $n(0)\zeta^d$, with $n(0)$ as the density of states at Fermi level and 
$\zeta$ as the localization length. Consequently, the typical energy separation between
such states  is $\Delta_{l}(E,Y)=(n(0)\zeta^d)^{-1}$. Similarly
the mean level spacing of states in the full length of the spectrum
is $\Delta(E,Y)=(n(0) L^d)^{-1}$ which gives $R_1=\Delta^{-1}=n(0) L^d$. 
For disordered systems, $\Delta_l$ can therefore be expressed in terms of the 
mean level density $R_1$: $\Delta_l=(L/\zeta)^d R_1^{-1}$.
\end{references}
\end{document}